\documentclass[10pt,twocolumn]{IEEEtran} %to change to the final version with two columns

\usepackage{theorem} \usepackage{cite} \usepackage{stfloats} \usepackage{epsfig} \usepackage{verbatim}
\usepackage{graphicx} \usepackage{amssymb} \usepackage{amsmath} \usepackage{color}
\usepackage{bm}
\usepackage{xcolor}
\usepackage{algorithm}
\usepackage{algorithmic}
\usepackage[caption=false,font=normalsize,labelfont=sf,textfont=sf]{subfig}
\usepackage{amsfonts}
\usepackage{makecell}

\theoremstyle{plain}
\newtheorem{theorem}{Theorem}
\theoremstyle{plain}

\theoremstyle{plain}

\theoremstyle{plain}

\renewcommand{\baselinestretch}{1}

\begin{document}

% Put a box at the end of the proof
\def\QEDclosed{\mbox{\rule[0pt]{1.3ex}{1.3ex}}}
\def\QEDopen{{\setlength{\fboxsep}{0pt}\setlength{\fboxrule}{0.2pt}\fbox{\rule[0pt]{0pt}{1.3ex}\rule[0pt]{1.3ex}{0pt}}}}
\def\QED{\QEDopen}
\def\proof{}
\def\endproof{\hspace*{\fill}~\QED\par\endtrivlist\unskip}

\title{An EEP-based Robust Beamforming Approach \\
for Superdirective Antenna Arrays \\
and Experimental Validations}
\author{Mengying~Gao,
    Haifan~Yin~\IEEEmembership{Member,~IEEE},
    and Liangcheng~Han

\thanks{Mengying Gao, Haifan Yin, and Liangcheng Han are with the School of Electronic Information and Communications, Huazhong University of Science and Technology, Wuhan 430074, China (e-mail: mengyinggao@hust.edu.cn; yin@hust.edu.cn; hanlc@hust.edu.cn). }%
\thanks{This work was supported by the National Natural Science Foundation of China under Grant 62071191. The corresponding author is Haifan Yin.}
}

\maketitle

\begin{abstract}
A superdirective antenna array has the potential to achieve an array gain proportional to the square of the number of antennas, making it of great value for future wireless communications. However, designing the superdirective beamformer while considering the complicated mutual-coupling effect is a practical challenge. Moreover, the superdirective antenna array is highly sensitive to excitation errors, especially when the number of antennas is large or the antenna spacing is very small, necessitating demanding and precise control over excitations. To address these problems, we first propose a novel superdirective beamforming approach based on the embedded element pattern (EEP), which contains the coupling information. The closed-form solution to the beamforming vector and the corresponding directivity factor are derived. This method relies on the beam coupling factors (BCFs) between the antennas, which are provided in closed form. To address the high sensitivity problem, we formulate a  constrained optimization problem and propose an EEP-aided orthogonal complement-based robust beamforming (EEP-OCRB) algorithm. Full-wave simulation results validate our proposed methods. Finally, we build a prototype of a 5-dipole superdirective antenna array and conduct real-world experiments. The measurement results demonstrate the realization of the superdirectivity with our EEP-based method, as well as the robustness of the proposed EEP-OCRB algorithm to excitation errors. 
\end{abstract}


%%%%%%%%%%%%%%%%%%%%%%%%%%%%%%%%%%%%%%%

%%%%%%%%%%%%%%%%%%%%%%%%%%%%%%%%%%%%%%% 

\section{Introduction}\label{sec:intro}
Massive multiple-input multiple-output (MIMO) is one of the key enablers of the fifth generation (5G) mobile communications. This technique utilizing arrays consisting of a large number of antennas at the base station leads to a large improvement in the spectral and energy efficiencies \cite{marzetta2010noncooperative,ngo2013energy}. However, with finite aperture of the array, the number of antennas is limited under the traditional constraint of half-wavelength antenna spacing, making it hard to push for higher spectral efficiency.  
Therefore, besides increasing the number of antennas, another way to improve the gain is being sought after, which is called superdirectivity. Recent work \cite{han2023superdirectivity} indicates that in a multi-user communication system, superdirectivity may greatly enhance the spectral efficiency, and meanwhile reduces the aperture of the antenna array. 

Superdirectivity states that an extraordinarily high directivity can be achieved by decreasing the element spacing in an antenna array, which challenges the common notion that an increase in the gain of an array must be accompanied by an increase in its aperture. 
Shelkunoff \cite{schelkunoff1943mathematical} first proved its feasibility in theory. Uzkov \cite{uzkov1946approach} was the first to consider the upper limit of directivity. He showed that for a linear endfire array of $M$ equispaced omnidirectional elements, the maximum directivity approaches $M^2$ as the spacing tends to $0$. Bloch \textit{et al.} \cite{bloch1953new} derived the closed-form maximum directivity and calculated the current distribution for linear arrays.

In reality, as the antenna spacing decreases, the coupling effect becomes stronger and can not be neglected in compact arrays. However, the works mentioned above do not take mutual-coupling between elements into consideration and can be categorized into the traditional superdirective beamforming method. 
Such traditional method uses the isolated element pattern (IEP) to represent each antenna element in the array pattern function and causes a decline of the directivity when the antenna spacing is small.
To address this issue, some researchers focus on decoupling techniques, such as designing decoupling networks \cite{andersen1976decoupling} or utilizing metamaterials \cite{bait2010electromagnetic}. Some researchers attempt to characterize the mutual-coupling. In \cite{clemente2015design,han2022coupling,han2023superdirective}, spherical wave expansion was used to describe the radiation field, which is a bit complicated when computing the spherical wave coefficients. The authors of \cite{marzetta2019super} addressed the problem from the point of view of circuit theory, utilizing the impedance matrix to describe the coupling information for omnidirectional elements. The calculation of impedances is complex and the extension to arrays of directive elements is so far unknown.

Another problem that hinders the practical use of superdirective antenna arrays is the sensitivity. A numerical analysis conducted by Yaru \cite{yaru1951note} showed that demanding and precise control over the magnitudes and phases of the excitations is required to attain the desired high directivity. Even small variations in the antenna system will cause significant fluctuations. Uzsoky and Solymar \cite{uzsoky1956theory} introduced the tolarance sensitivity $T$ and presented a mathematical procedure for determining the maximum directivity for a given $T$. Gilbert and Morgan \cite{gilbert1955optimum} defined the sensitivity factor $K$ which is related to the electrical and mechanical tolerance of an antenna. The authors of \cite{newman1978superdirective} presented numerical data that illustrated the trade-offs between directive gain, efficiency, bandwidth, farfield patterns and $K$. However, these analyses are all in the framework of the traditional superdirective beamforming since they ignore mutual-coupling. In \cite{ruze1966antenna} and \cite{trucco2022statistics}, statistical analysis of random errors was conducted. 

In this paper, we develop novel superdirective beamforming appraoches that take into account the mutual-coupling effect between antennas. To achieve superdirectivity in reality, we introduce the embedded element pattern (EEP) to characterize the coupling effect in the antenna array concisely.
The modified expression of the directivity factor involves the beam coupling factors (BCFs), which are the key parameters in our proposed method. We show that, under actual electromagnetic conditions, they can be either computed through numerical integration with field samples or derived from the scattering parameters or the array impedances.

To gain an insight into the sensitivity problem, we define a metric that measures the susceptibility of the antenna array to excitation variations. Then, on the basis of evaluating the sensitivity and the EEP, we propose a robust superdirective beamforming method based on orthogonal complement. 

Simulations and experiments with uniform linear dipole arrays have validated our proposed approaches.

Our contributions are as follows: 
\begin{itemize}
  \item We propose an EEP-based superdirective beamforming (EEPB) method, which takes the mutual-coupling into consideration. The closed-form solutions of the optimum excitation vector and the maximum directivity factor are derived. We show that, in a compact array, this approach significantly alleviates the problem of directivity deterioration caused by coupling and achieves ``superdirectivity".
  \item The proposed superdirective beamforming method entails certain BCFs. Computing the BCFs through numerical integration usually requires numerous field points to ensure accurate results. Therefore, we derive the matrix of the BCFs from the array scattering matrix or the array impedance matrix directly in a lossless antenna array, circumventing the complexity of numerical integration.
  \item To constrain the sensitivity of the antenna array system, a valid metric should be defined first. In this context, we introduce the normalized variance of the array pattern to quantify the array sensitivity for beamforming. By leveraging the central limit theorem, we derive the closed-form expression of the normalized variance.
  \item To combat the high sensitivity problem, we propose an EEP-aided orthogonal complement-based robust superdirective beamforming (EEP-OCRB) algorithm that offers a trade-off between the directivity and the sensitivity. The closed-form solution to the excitations is derived. 
  With the same extent of excitation variations, our proposed approach exhibit less directivity fluctuations.
  \item We construct a prototype of superdirective antenna array and test it in the microwave anechoic chamber to validate the performance and the robustness of our proposed approaches. The array consists of 5 dipole antennas operating at 1.6 GHz. We validate the superdirectivity of the proposed EEPB method and verify the robustness of the proposed EEP-OCRB algorithm. To the best of our knowledge, it is the first time to realize the robust superdirectivity in a five-antenna array in the real-world experiments.
\end{itemize}

This paper is organized as follows: In Sec. \ref{sec:SuperdirBeamformingMethods} we recall the traditional superdirective beamforming method and then propose a novel superdirective beamforming method based on the EEP. Sec. \ref{sec:SensitivityEvaluation} evaluates the antenna array sensitivity. An algorithm of the robust superdirective beamforming is presented in Sec. \ref{sec:RobustSuperdirBeamforming}. We provide simulation results in Sec. \ref{sec:SimulationResults} and experimental results in Sec. \ref{sec:ExperimentalResults}, respectively. Finally, conclusions are drawn in Sec. \ref{sec:conclusions}.

\emph{Notation:} Boldface letters denote matrices and vectors. Specifically, ${\mathbf{I}}_M$ denotes the $M \times M$ identity matrix. The superscripts ${(\cdot)^T}$, ${(\cdot)^*}$, and ${(\cdot)^H}$ stand for the transpose, conjugate, and conjugate transpose, respectively. $\left|\cdot\right|$ denotes the absolute value. $\mathbb{E}\{\cdot\}$ indicates the expectation. $\Re \left\{ \cdot \right\}$ stands for the real part. ${\mathop{\rm diag}\nolimits} \{ {\bf{a_1,\cdots,a_N}}\}$ refers to a diagonal matrix or a block diagonal matrix with $\bf{a_1,\cdots,a_N}$ at the main diagonal. ${z} \sim {\cal N}\left( {0,{\sigma}^2 } \right)$ describes a real-valued Gaussian random variable $z$ with zero mean and variance ${\sigma}^2$. $\mathbb{C}^M$ is a vector space with $M$ elements. ${\text{det}}\left\{{\bf{A}}\right\}$ refers to the determinant of the square matrix ${\bf{A}}$. $span\left\{\cdots\right\}$ represents the set of all linear combinations of a given set of vectors. The inner product of two vectors $\bf{x}$ and $\bf{y}$ is denoted by $\left( {{\bf{x}},{\bf{y}}} \right)$. The symbol $\triangleq$ is used for definition.

\section{Superdirective Beamforming Methods}\label{sec:SuperdirBeamformingMethods}
\subsection{Traditional Superdirective Beamforming based on the IEP}\label{sec:IEP}
In the traditional superdirective beamforming method, the pattern multiplication rule \cite{balanis2016antenna} is utilized for arrays of identical elements, which ignores the mutual-coupling and the accompanying pattern distortion. When the elements are closely spaced and strongly coupled, this method leads to array directivity deterioration.

Consider an array of $M$ identical elements. The traditional method describes its pattern as follows:
\begin{equation}\label{Eq:IEP_pattern}
{\bf{F}}\left( {\bf{u}} \right) = \sum\limits_{i = 1}^M {{a_i}\exp \left( {ik{{\bf{r}}_i}{\bf{u}}} \right){\bf{f}}\left( {\bf{u}} \right)},
\end{equation}
where ${a}_{i}$ is the excitation coefficient on the $i$-th element, which is a complex value. $\bf{a}$ denotes the excitation vector:
\begin{equation}\label{Eq:a}
{\bf{a}} = \begin{bmatrix}
a_1  & a_2  & \cdots  & a_M
\end{bmatrix}^T.
\end{equation}
$k = \frac{{2\pi }}{\lambda }$ is the wavenumber, while $\lambda$ denotes the wavelength in the medium. ${{\bf{r}}_i}$ is the position vector of the $i$-th element with a specified origin as reference, and ${\bf{u}}$ is the unit vector representing the direction of the field point. ${\bf{f}}\left( {\bf{u}} \right)$ is the radiation pattern of the antenna isolated in unbounded free space, taking the center of the individual element as the origin, which is called the isolated element pattern (IEP). Suppose the direction of the maximum radiation intensity is ${{\bf{u}}_0}$, then the directivity factor of the antenna array is given by \cite{balanis2016antenna}: 
\begin{equation}\label{Eq:D}
D\left( {\bf{a}} \right) = \frac{{{{\left| {{\bf{F}}\left( {{{\bf{u}}_0}} \right)} \right|}^2}}}{{\frac{1}{{4\pi }}\int_S {{{\left| {{\bf{F}}\left( {\bf{u}} \right)} \right|}^2}d\Omega } }},
\end{equation}
where $\Omega$ is the solid angle, and $S$ is the surface of the unit sphere. ${\bf{a}}$ can be specially chosen to maximize this pattern function. Obviously, Eq. (\ref{Eq:D}) is written in the form of Rayleigh quotient. By applying Rayleigh-Ritz theorem, one may readily derive the optimum ${\bf{a}}$ and the maximum $D$ \cite{han2022coupling,han2023superdirective}.

However, it should be noted that the method employs an approximation, namely representing the radiation patterns of the identical elements embedded in the array using the IEP, which serves as the foundation of the method. Although using this approach in widely spaced and weakly coupling situation attains much higher directivity than other conventional ways, it performs worse and worse as the antenna spacing decreases and coupling effect enhances. Strong coupling makes the actual pattern of the embedded elements deviate a lot from the IEP, rendering the aforementioned approximation inadequate. In that case, ${\bf{a}}$ obtained from maximizing (\ref{Eq:D}) is not the optimum excitation vector. As a result, the directivity declines rapidly, indicating that the IEP method fails to achieve superdirectivity when the elements are too close together.

\subsection{Proposed Superdirective Beamforming based on the EEP}\label{sec:EEP}
To achieve superdirectivity, especially in closely spaced and strongly coupling arrays, we propose an embedded element pattern-based superdirective beamforming (EEPB) method. In the presence of another object in the vicinity of the antenna, its boundary conditions change and new currents are induced, which impacts the radiation pattern. For better description of the radiation pattern, we discard the pattern multiplication rule and reformulate the array pattern as:
\begin{equation}\label{Eq:EEP_pattern}
{\bf{F}}\left( {\bf{u}} \right) = \sum\limits_{i = 1}^M {{a_i}{{\bf{f}}_i}\left( {\bf{u}} \right)},
\end{equation}
where ${{\bf{f}}_i}\left( {\bf{u}} \right)$ is the radiation pattern of the $i$-th element when it is the only excited element in an array and others are terminated, which is called the embedded element pattern (EEP) \cite{craeye2011review}. In electromagnetic simulations, it can be exported directly from the simulation software. To obtain it in the experiment, we need the amplitude of the electric field and the corresponding phase in different solid angles. 
Note that no matter for ${\bf{F}}\left( {\bf{u}} \right)$ or any ${{\bf{f}}_i}\left( {\bf{u}} \right)$, a common point is selected as the origin and used for phase reference.

Substituting (\ref{Eq:EEP_pattern}) into (\ref{Eq:D}), we obtain the expression of the array directivity:
\begin{equation}\label{Eq:EEP_D}
D\left( {\bf{a}} \right) = \frac{{{{\left| {{{\bf{a}}^T}{{\bf{v}}_0}} \right|}^2}}}{{{{\bf{a}}^T}{\bf{B}}{{\bf{a}}^*}}},
\end{equation}
where ${{\bf{v}}_0}$ is an $M \times 1$ vector defined as
\begin{equation}\label{Eq:EEP_F0}
{\bf{v}}_0 = \begin{bmatrix}
{{\bf{f}}_1}\left( {{{\bf{u}}_0}} \right)  & {{\bf{f}}_2}\left( {{{\bf{u}}_0}} \right)  & \cdots  & {{\bf{f}}_M}\left( {{{\bf{u}}_0}} \right)
\end{bmatrix}^T.
\end{equation}
${\bf{B}}$ is an $M \times M$ Hermitian matrix of the BCFs, with its entries defined as:
\begin{equation}\label{Eq:EEP_BCF}
{B_{ij}} = \frac{1}{{4\pi }}\int_S  {{{\bf{f}}_i}\left( {\bf{u}} \right){{\bf{f}}_j}^H\left( {\bf{u}} \right)d\Omega },
\end{equation}
which is located in row $i$, column $j$ of ${\bf{B}}$ and is the BCF between elements $i$ and $j$ \cite{craeye2011review}. 

In the following theorem, we show the closed-form solution to the problem of maximizing the array directivity defined in (\ref{Eq:EEP_D}) with respect to ${\bf{a}}$.
\begin{theorem}\label{theoRayleighRitz}
The directivity factor (\ref{Eq:EEP_D}) reaches its maximum:
\begin{equation}\label{Eq:EEP_D0}
{D_0} = {{\bf{v}}_0}^H{{\bf{B}}^{ - 1}}{{\bf{v}}_0},
\end{equation}
when the array is excited by the excitation vector ${\bf{a}}_0^*$:
\begin{equation}\label{Eq:EEP_a0}
{{\bf{a}}_0}^* = \beta{{\bf{B}}^{ - 1}}{{\bf{v}}_0},
\end{equation}
where $\beta$ is a constant depending on the power constraint.
\end{theorem}
\begin{proof}
\quad \emph{Proof:}
See Appendix \ref{proof:theoRayleighRitz}.
\end{proof}

It is worthy to point out that no restrictions on array configuration, antenna types nor radiation direction have been imposed on the derivation of this EEPB method, hence it can be generalized to arbitrary arrays.

Eqs. (\ref{Eq:EEP_D0}) and (\ref{Eq:EEP_a0}) reveal that the key of the EEPB method is to obtain ${{\bf{v}}_0}$ and the BCFs. According to (\ref{Eq:EEP_F0}), ${{\bf{v}}_0}$ only requires the information about the radiation field in ${{\bf{u}}_0}$ direction and is easy to obtain. The main difficulty lies in the acquisition of the BCFs. For antenna arrays whose element patterns can not be approximated by a ${\cos ^q}\theta $ variation, one has to resort to numerical integration to obtain the BCFs. The numerical computation of (\ref{Eq:EEP_BCF}) calls for a large number of field points to yield accurate results in complex cases, which is cumbersome in practice. Thus, we introduce some simpler approaches below to calculate the BCFs.

For lossless antenna arrays, the BCFs can be directly obtained from the array scattering matrix or array impedance matrix, both of which can be readily measured using vector network analyzers (VNA). This avoids the numerical integration. The relationships between these quantities are based on energy conservation, and are derived in detail in Appendix \ref{proof:BCF}.

Fig. \ref{fig:ThevininConfig} shows the general configuration of an M-antenna array, with Thevinin equivalent representations of the generators.
\begin{figure}[b]
  \centering
  \includegraphics[width=3.2in]{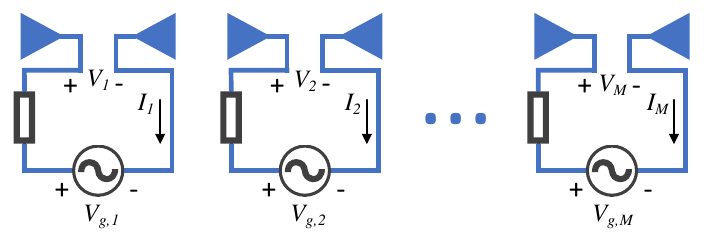}\\
  \caption{The general configuration of an M-antenna array, with Thevinin equivalent representations of the generators.} \label{fig:ThevininConfig}
\end{figure}
For the $i$-th port, ${V_{g,i}}$ represents the voltage source of the Thevinin equivalent circuit, ${V_i}$ is the total voltage at the port, and ${I_i}$ is the total current flows through the transmission line. Inspired by \cite{stein1962cross}, a relationship between the BCFs and the scattering parameters in our case can be derived. When all the ports are terminated with the same reference impedance $Z_0$, which is usually real, $\bf{B}$ can be expressed in terms of the array scattering matrix $\bf{S}$ as:
\begin{equation}\label{Eq:B(S)}
{\bf{B}} = \frac{\eta }{{16\pi \Re \left\{ {{Z_0}} \right\}}}\left( {{{\mathbf{I}}_M} - {{\bf{S}}^T}{{\bf{S}}^*}} \right),
\end{equation}
where $\eta$ is the free space impedance.
When the reference impedances at each of the ports are different and have complex values, the relationship above need to be generalized. Thus, the generalized scattering parameters \cite{collin2007foundations}, denoted by ${S}_G$ here, are introduced:
\begin{equation}\label{Eq:GeneralizedS}
{\bf{S}}_G = {\left( {{{\bf{G}}^*}} \right)^{ - 1}}\left( {{\bf{S}} - {{\bf{\Gamma }}^*}} \right){\left( {{{\mathbf{I}}_M} - {\bf{\Gamma S}}} \right)^{ - 1}}{\bf{G}},
\end{equation}
where $\bf{D}$ is a diagonal matrix with elements
\begin{align}
{G_{ii}} &= {\left| {1 - {\Gamma _i}^*} \right|^{ - 1}}\left( {1 - {\Gamma _i}} \right)\sqrt {1 - {{\left| {{\Gamma _i}} \right|}^2}}\label{Eq:GeneralizedS_Gii}\\
{\Gamma _i} &= \frac{{{Z_{0,i}} - {Z_0}}}{{{Z_{0,i}} - {Z_0}}}, i = 1,2, \cdots ,M, \label{Eq:GeneralizedS_Gammai}
\end{align}
and $\bf{\Gamma}$ is also a diagonal matrix with elements $\Gamma _i$. ${Z_{0,i}}$ is the reference impedance at the $i$-th port. Then an extended version of (\ref{Eq:B(S)}) is derived:
\begin{equation}\label{Eq:B(GeneralizedS)}
{\bf{B}} = \frac{\eta }{{16\pi }}\Re {\left\{ {{{\bf{Z}}_0}} \right\}^{ - \frac{1}{2}}}\left( {{\mathbf{I}}_M} - {{{\bf{S}}_G}^T}{{{\bf{S}}_G}^*} \right)\Re {\left\{ {{{\bf{Z}}_0}} \right\}^{ - \frac{1}{2}}},
\end{equation}
where
\begin{equation}\label{Eq:Z0matrix}
{{\bf{Z}}_0} = {\mathop{\rm diag}\nolimits} \{ {Z_{0,1}, Z_{0,2}, \cdots, Z_{0,M}}\}.
\end{equation}
A very good agreement has been verified when comparing the BCFs computed with ${\bf{S}}_G$ and the BCFs obtained from numerical integration of the EEPs \cite{craeye2011review}. 

Since an antenna array can be represented by an equivalent electrical network \cite{ivrlavc2010toward}, there is also an expression of $\bf{B}$ in terms of the $M \times M$ array impedance matrix $\bf{Z}$:
\begin{equation}\label{Eq:B(Z)}
{\bf{B}} = \frac{\eta }{{4\pi }}{\left( {{{\left( {{\bf{Z}} + {{\bf{Z}}_0}} \right)}^{ - 1}}} \right)^T}\Re \left\{ {\bf{Z}} \right\}{\left( {{{\left( {{\bf{Z}} + {{\bf{Z}}_0}} \right)}^{ - 1}}} \right)^*}.
\end{equation}
Note that this relationship holds regardless of the reference impedances to which the ports are connected.
The derivations of (\ref{Eq:B(S)}), (\ref{Eq:B(GeneralizedS)}) and (\ref{Eq:B(Z)}) are all presented in Appendix \ref{proof:BCF}.

The above relationships are derived solely by utilizing the energy conservation conditions. Although realizing a completely lossless antenna array in reality is challenging, such relations still provide a useful insight for calculating the BCFs in a simpler way. Further research is required to develop practical modifications for these relationships that can be applied to actual arrays in practice.

The EEPB method has provided a viable implementation approach for achieving superdirectivity in the presence of strong coupling. However, to apply it in practical scenarios, the sensitivity is also a crucial factor to consider. Next we will evaluate the sensitivity of different beamforming methods.

\section{Evaluation of Beamforming Sensitivity}\label{sec:SensitivityEvaluation}
\subsection{Sensitivity Analysis}\label{sec:SensitivityAnalysis}
In spite of extraordinarily high directivity, superdirective beamforming is very sensitive to perturbations. Slight errors on excitations may lead to significant decline in array directivity \cite{yaru1951note,newman1978superdirective}. In this section, we conduct a sensitivity analysis on antenna arrays to examine the impact of errors, specifically the degree of directivity fluctuations. The errors on the excitations of antenna arrays can be categorized into systematic errors and random errors. The former, once known or measured, can be dealt with by a suite of standard approaches \cite{ruze1966antenna}. The latter are trickier. Thus, we focus on the random errors on excitations.

\begin{figure*}[t]
\centering
\subfloat[]{\includegraphics[width=2.3in]{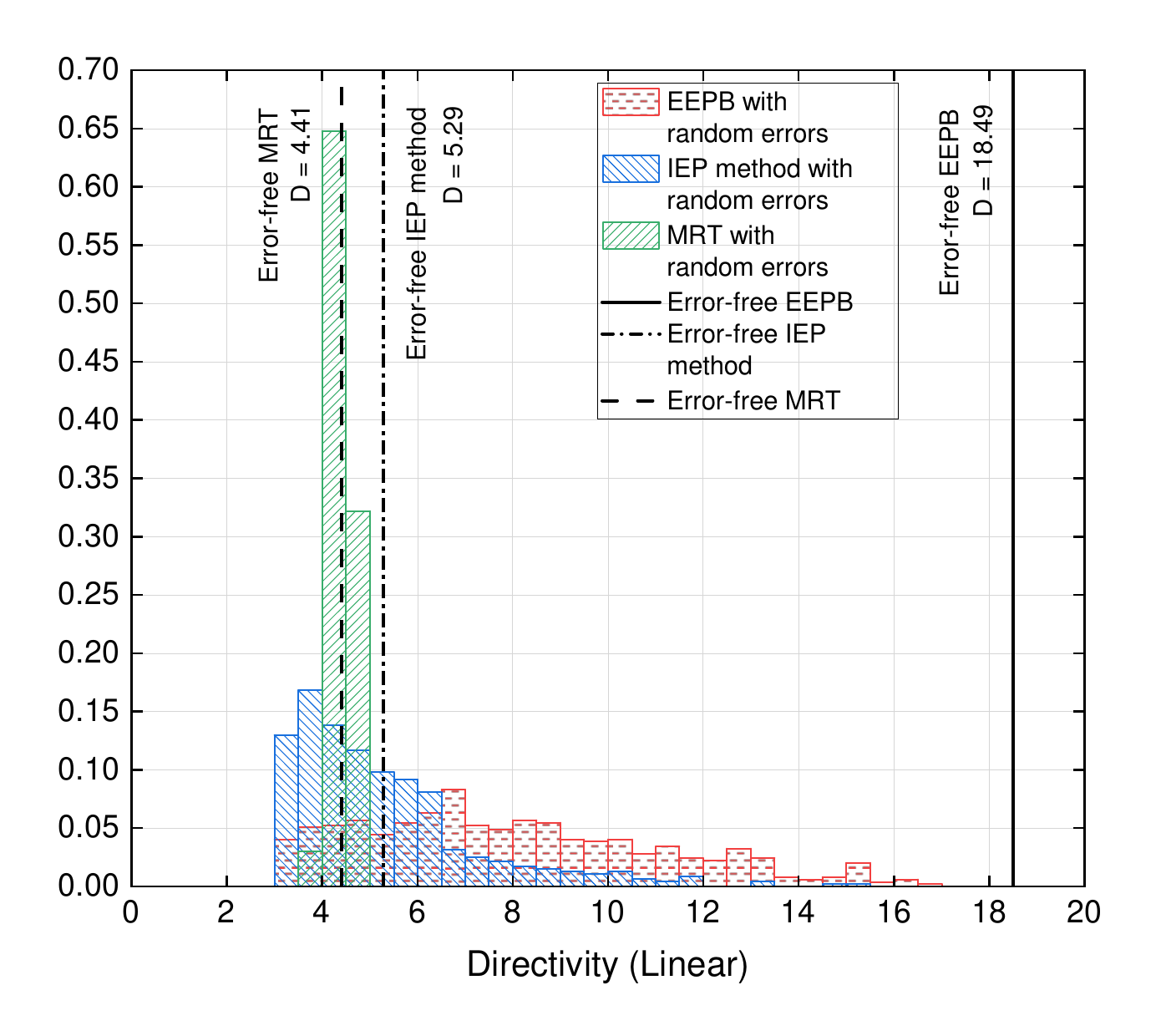}%
\label{fig:M=4MCSim_0.10lambda}}
\hfil
\subfloat[]{\includegraphics[width=2.3in]{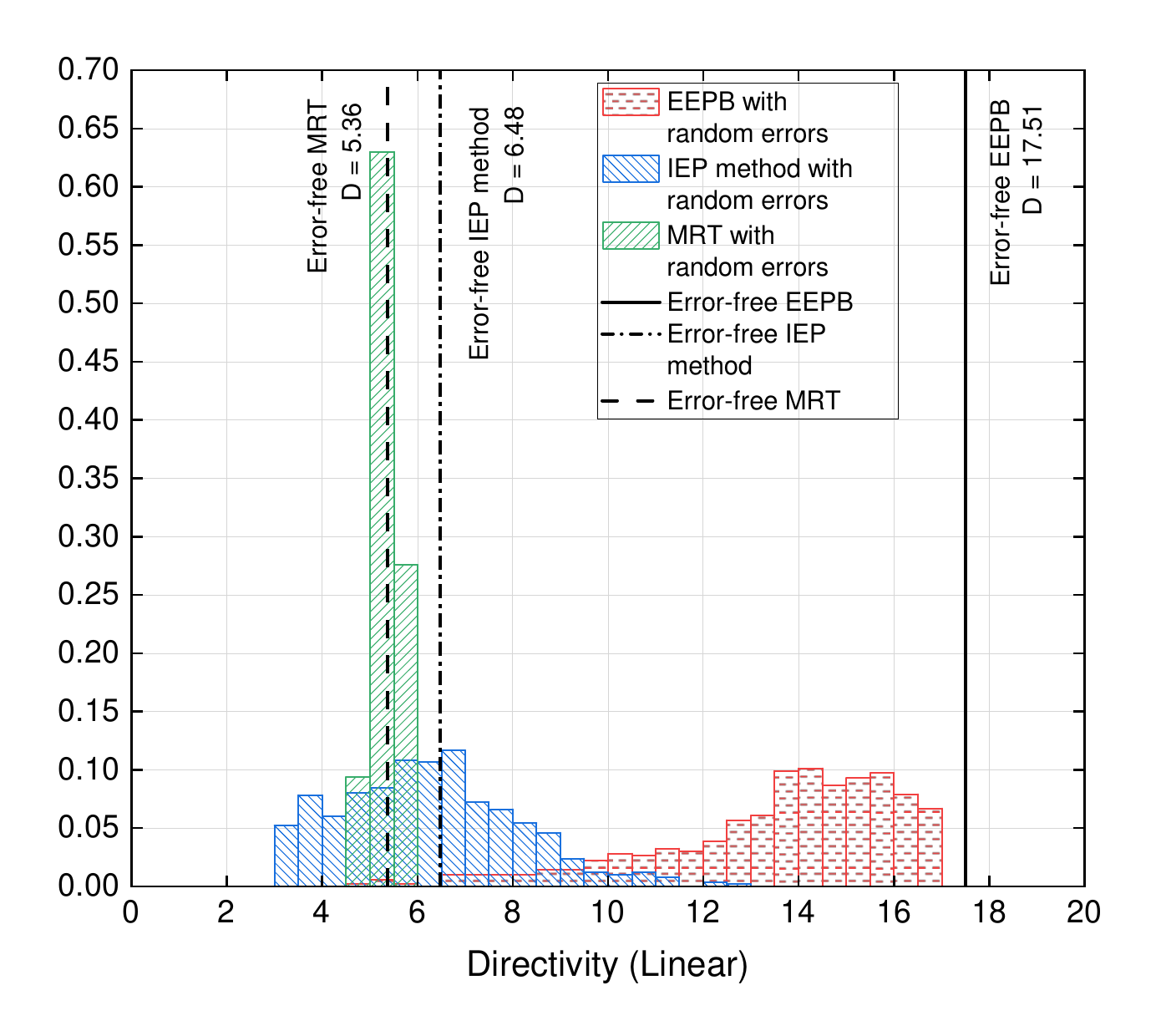}%
\label{fig:M=4MCSim_0.15lambda}}
\hfil
\subfloat[]{\includegraphics[width=2.3in]{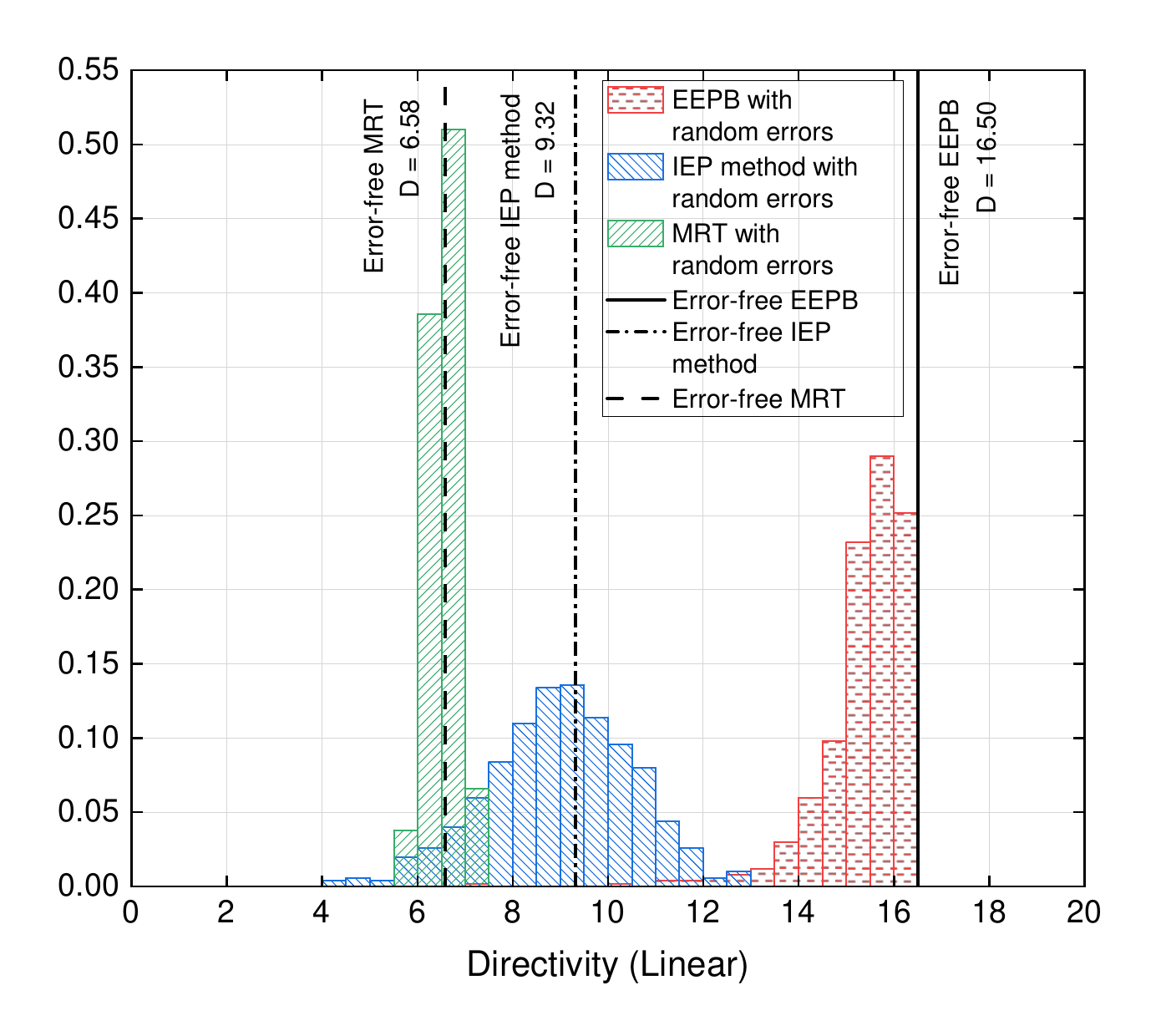}%
\label{fig:M=4MCSim_0.20lambda}}
\caption{The directivity histograms of different beamforming methods in Monte Carlo simulations at different spacings---(a) $0.10\lambda$, (b) $0.15\lambda$ and (c) $0.20\lambda$---along with the error-free directivity factors.}
\label{fig:M=4MCSim}
\end{figure*}

We express the actual excitation on the $i$-th element as ${a_i}\left( {1 + {\alpha _i}} \right){e^{j{\delta _i}}}$, where $\alpha _i$ is a scalar denoting the relative amplitude error and $\delta _i$ is the phase error. ${\alpha _i}\left( {i = 1,2, \cdots , M} \right)$ and ${\delta _i}\left( {i = 1,2, \cdots , M} \right)$ are all independently and identically Gaussian distributed \cite{ruze1966antenna,gilbert1955optimum}, i.e.,
\begin{equation}\label{Eq:NormalDistribution}
{\alpha _i} \sim {\cal N}\left( {0,\overline {{\alpha ^2}} } \right), {\delta _i} \sim {\cal N}\left( {0,\overline {{\delta ^2}} } \right), i = 1,2, \cdots ,M,
\end{equation}
where $\overline {{\alpha ^2}}$ and $\overline {{\delta ^2}}$ are the variances of $\alpha _i$ and $\delta _i$, respectively.
When the nominal excitation vector $\bf{a}$ is known and the distributions of errors are given, Monte Carlo simulations are carried out, repeating $N$ times for a fixed array. One simulation generates an $M\times1$ excitation vector involving random errors sampled from known distributions:
\begin{equation*}
\begin{bmatrix}
{a_1}\left( {1 + {\alpha _1}} \right){e^{j{\delta _1}}} \\
{a_2}\left( {1 + {\alpha _2}} \right){e^{j{\delta _2}}} \\
\vdots \\
{a_M}\left( {1 + {\alpha _M}} \right){e^{j{\delta _M}}}
\end{bmatrix},
\end{equation*}
with which the antenna array is stimulated. We apply three beamforming methods---the EEPB method, the traditional IEP method, and maximum ratio transmission (MRT)---to the antenna array in Fig. \ref{fig:SchematicViewOfDipoleArray} in full-wave simulations. Assuming $\sqrt{\overline {{\alpha ^2}}} = 0.05$, $\sqrt{\overline {{\delta ^2}}} = {5^ \circ }$, and $N=500$, the histograms of the directivity factors obtained from Monte Carlo simulations are presented in Fig. \ref{fig:M=4MCSim}.
Besides, for different beamforming methods, the desired directivity factors, which are achieved when the excitation coefficients are error-free, are marked as references. It is clear that the same extent of excitation variations leads to more significant directivity fluctuations in the EEPB method and the IEP method than in MRT beamforming, which means the superdirective beamforming methods are more sensitive to errors. The smaller antenna spacing leads to the higher sensitivity.

The histogram may not provide a clear indication of which beamforming method is more sensitive. Therefore, we introduce a metric that offers a numerical representation of the degree of directivity fluctuations. When the antenna array is excited without errors, we can obtain the desired radiation pattern ${{\bf{F}}_0}\left( {\bf{u}} \right)$, which has a standard error-free directivity $D_0$. $N$ Monte Carlo simulations for a fixed antenna array yield a distribution of directivity, containing $N$ directivity factors. The deviated directivity obtained from the $i$-th Monte Carlo simulation is ${D_i}\left( {i = 1,2, \cdots ,N} \right)$. We define:
\begin{equation}\label{Eq:H}
\mathcal{H} \triangleq \frac{{\sum\limits_{i = 1}^N {{{\left| {{D_i} - {D_0}} \right|}^2}} }}{N}.
\end{equation}
It indicates the degree to which the directivity distribution deviates from $D_0$. $\mathcal{H}$ is a depiction of directivity fluctuations based on the data obtained from a number of samplings. The samples can be taken from either simulations or experiments.

\subsection{Quantification of Sensitivity}\label{sec:NormalizedVariance}
Although $\mathcal{H}$ is a measurement of the sensitivity, it requires prior knowledge of the excitation vector ${\bf{a}}$ and its accuracy depends on the number of $N$. Conducting a large number of Monte Carlo simulations is laborious; hence, we need another metric to describe the sensitivity of an excited antenna array quantitatively. 
The normalized variance of the array pattern function ${\bf{F}}\left( {\bf{u}} \right)$ is introduced to measure the sensitivity. It provides an exact numerical value to intuitively reflect how susceptible the array pattern is to excitation perturbations in different beamforming methods, without the requirement for Monte Carlo simulations. In this paper, we denote the normalized variance of the array pattern by $\Xi$. It is defined as follows:
\begin{equation}\label{Eq:NormalizedVariancedefinition}
\begin{split}
\Xi &\triangleq \frac{{\mathbb{E}\left\{ {{{\left| {{\bf{F}}\left( {\bf{u}} \right) - \mathbb{E}\left\{ {{\bf{F}}\left( {\bf{u}} \right)} \right\}} \right|}^2}} \right\}}}{{{{\left| {\mathbb{E}\left\{ {{\bf{F}}\left( {\bf{u}} \right)} \right\}} \right|}^2}}} \\
&= \frac{{\mathbb{E}\left\{ {{{\left| {{\bf{F}}\left( {\bf{u}} \right)} \right|}^2}} \right\} - {{\left| {\mathbb{E}\left\{ {{\bf{F}}\left( {\bf{u}} \right)} \right\}} \right|}^2}}}{{{{\left| {\mathbb{E}\left\{ {{\bf{F}}\left( {\bf{u}} \right)} \right\}} \right|}^2}}}.
\end{split}
\end{equation}

Then we discuss the statistical properties of the random errors on excitations:
\begin{itemize}
\item First, for an $M$-element antenna array where the amplitude and the phase on each element are controlled separately, the errors can be represented by $2M$ statistically independent random variables, $M$ for random errors on amplitudes and $M$ for phases. 
\item Second, usually in practice, all the antennas are controlled by identical amplitude or phase controllers. Therefore, the $M$ random variables representing amplitude errors are independent and identically distributed, and the $M$ variables indicating phase errors are also independent and identically distributed. 
\item Third, the amplitude or phase error on a certain excitation is the sum of numerous random and generally independent contributions \cite{ruze1966antenna}. According to the central limit theorem of statistical theory, the error will be distributed in an asymptotically Gaussian manner with a mean $0$ and a variance $\sigma^{2}$ determined by the specific case. 
\end{itemize}
These three properties justify the independent and identically distributed Gaussian distribution of random errors on excitations, as described in (\ref{Eq:NormalDistribution}).

The following theorem presents the closed-form expression for $\Xi$, i.e., the normalized variance of the array pattern.

\begin{theorem}\label{theoNormalizedVariance}
Based on the three statistical properties listed above, the closed-form expression of $\Xi$ for an $M$-element array is given by:
\begin{equation}\label{Eq:NormalizedVariance}
\Xi = \frac{{{{\bf{a}}^T}{{\bf{D}}_{{f_0}}}{{\bf{a}}^*}}}{{{{\left| {{{\bf{a}}^T}{{\bf{v}}_0}} \right|}^2}}},
\end{equation}
where
\begin{equation}
{{\bf{D}}_{{f_0}}} = {\mathop{\rm diag}\nolimits} \{ {{{\left| {{{\bf{f}}_1}\left( {{{\bf{u}}_0}} \right)} \right|}^2},{{\left| {{{\bf{f}}_2}\left( {{{\bf{u}}_0}} \right)} \right|}^2},\cdots,{{\left| {{{\bf{f}}_M}\left( {{{\bf{u}}_0}} \right)} \right|}^2}}\}.
\end{equation}
\end{theorem}
\begin{proof}
\quad \emph{Proof:}
See Appendix \ref{proof:theoNormalizedVariance}.
\end{proof}

Note that unit scaling has been done during the derivation of (\ref{Eq:NormalizedVariance}), hence $\Xi$ can be regarded as the array pattern fluctuation caused by unit excitation variation. Theorem \ref{theoNormalizedVariance} reveals that $\Xi$, the normalized variance of the array pattern, has a similar form to the directivity $D$ and is also a ratio of two Hermitian quadratic forms. It can be seen that the value of $\Xi$ depends on the excitation coefficients, the topology of the array, and the types of the antenna elements.

Making use of Theorem \ref{theoRayleighRitz}, we easily derive from (\ref{Eq:NormalizedVariance}) that
\begin{equation}\label{Eq:NormalizedVariancerange}
\Xi \ge \frac{1}{{{{\bf{v}}_0}^H{{\bf{D}}_{{f_0}}}^{ - 1}{{\bf{v}}_0}}} = \frac{1}{M},
\end{equation}
and $\Xi$ reaches its minimum when
\begin{equation}\label{Eq:ConditionOfNormalizedVariancemin}
{\bf{a}} \propto \begin{bmatrix}
\frac{1}{{{\bf{f}}_1}\left( {{{\bf{u}}_0}} \right)}  & \frac{1}{{{\bf{f}}_2}\left( {{{\bf{u}}_0}} \right)}  & \cdots  & \frac{1}{{{\bf{f}}_M}\left( {{{\bf{u}}_0}} \right)}
\end{bmatrix}^T.
\end{equation}
In the following sections, Eq. (\ref{Eq:NormalizedVariance}) is used to measure the sensitivity of a fixed array stimulated with given excitations. A smaller value of $\Xi$ signifies a more robust beamforming method. In brief, $\Xi$ is a theoretical characterization of the array pattern fluctuations based on statistical theory.

To facilitate the design of the beamformer, we use $\Xi$ to evaluate the sensitivity instead of $\mathcal{H}$, circumventing the necessity for prior knowledge of ${\bf{a}}$ and the sampling. The validity of the metric $\Xi$ is supported by the positive correlation between $\mathcal{H}$ and $\Xi$, which is easily verified by simulations. 

Unstable superdirectivity hinders its practical applications. In the following section, we will propose a robust superdirective beamforming approach to address the sensitivity problem.

\section{Robust Superdirective Beamforming based on Orthogonal Complement}\label{sec:RobustSuperdirBeamforming}
The EEPB method yields exceedingly high directivity, yet is also corrupted by high sensitivity.
In fact, designers prefer adequate directivity that is resistant to excitation errors rather than unstable and excessively high directivity. Hence, a trade-off between the directivity and the sensitivity need to be striven in practice. Based on the EEPB method, we formulate the problem of robust superdirective beamforming below.

Our goal is to determine the excitation vector for maximum directivity under a constraint on the sensitivity. It can be written as a constrained optimization problem \cite{cheng1971optimization}:
\begin{equation}\label{Eq:ConstrainedOpt_D}
\mathop {\max }\limits_{\bf{a}} D\left( {\bf{a}} \right) = \frac{{{{\left| {{{\bf{a}}^T}{{\bf{v}}_0}} \right|}^2}}}{{{{\bf{a}}^T}{\bf{B}}{{\bf{a}}^*}}},
\end{equation}
subject to the constraint
\begin{equation}\label{Eq:ConstrainedOpt_NormalizedVariance}
\Xi = \xi.
\end{equation}

An EEP-aided orthgonal complement-based robust superdirective beamforming (EEP-OCRB) method is proposed to solve this optimization problem. In the following theorem, we provide the closed-form solution to the constrained optimization problem above.
\begin{theorem}\label{theoConstrainedOpt_a}
Under the constraint of $\Xi = \xi$, the array directivity factor reaches its maximum when the array is excited by the beamforming vector that satisfies the condition:
\begin{equation}\label{Eq:ConstrainedOpt_a}
{{\bf{a}}^*} = q{{\bf{K}}^{ - 1}}{{\bf{v}}_0},
\end{equation}
where
\begin{align}
q &= \frac{{{{\bf{a}}^T}{\bf{B}}{{\bf{a}}^*}}}{{{{\left| {{{\bf{a}}^T}{{\bf{v}}_0}} \right|}^2}}}{{\bf{v}}_0}^H{{\bf{a}}^*}\label{Eq:ConstrainedOpt_a_q}\\
{\bf{K}} &= {\bf{B}} + p \cdot \xi \cdot {{\bf{v}}_0}{{\bf{v}}_0}^H - p \cdot {{\bf{D}}_{{f_0}}}\label{Eq:ConstrainedOpt_a_K}\\
p &= \Lambda  \cdot \xi \cdot \frac{{{{\left( {{{\bf{a}}^T}{\bf{B}}{{\bf{a}}^*}} \right)}^2}}}{{{{\bf{a}}^T}{{\bf{D}}_{{f_0}}}{{\bf{a}}^*}{{\left| {{{\bf{a}}^T}{{\bf{v}}_0}} \right|}^2}}},\label{Eq:ConstrainedOpt_a_p}
\end{align}
and $\Lambda$ is a scalar multiplier introduced in the method of Lagrange multipliers.
\end{theorem}
\begin{proof}
\quad \emph{Proof:}
See Appendix \ref{proof:theoConstrainedOpt_a}.
\end{proof}

Theorem \ref{theoConstrainedOpt_a} indicates that the optimum beamforming vector under a constraint on $\Xi$ is a generalized version of (\ref{Eq:EEP_a0}). Specifically, when $p=0$, Eq. (\ref{Eq:ConstrainedOpt_a}) is reduced to the solution of the unconstrained situation which is described in (\ref{Eq:EEP_a0}).

Since only the relative magnitudes and relative phases of $a_1$, $\cdots$, $a_M$ are of interest, the scalar factor $q$ can be ignored. Thus, the only unknown in (\ref{Eq:ConstrainedOpt_a}) is $p$, which is proportional to the Largrange multiplier $\Lambda$ and the normalized variance $\Xi$. Inserting (\ref{Eq:ConstrainedOpt_a}) into (\ref{Eq:ConstrainedOpt_NormalizedVariance}), the following relation containing $p$ holds true:
\begin{equation}\label{Eq:ConstrainedOpt_EqOfp}
{{\bf{v}}_0}^H{{\bf{K}}^{ - 1}}\left( {\xi \cdot {{\bf{v}}_0}{{\bf{v}}_0}^H - {{\bf{D}}_{{f_0}}}} \right){{\bf{K}}^{ - 1}}{{\bf{v}}_0} = 0.
\end{equation}

The theorem below provides the optimum value of the parameter $p$ that satisfies the given relationship.
\begin{theorem}\label{theoConstrainedOpt_p}
Solving (\ref{Eq:ConstrainedOpt_EqOfp}) is equivalent to solving the equation involving $p$ below:
\begin{equation}\label{Eq:ConstrainedOpt_detW}
{\text{det}}\left\{{\bf{W}}\right\} = {\text{det}} \begin{bmatrix}
{{\bf{v}}_0} & {{\bf{w}}_2} & {{\bf{w}}_3} & \cdots &{{\bf{w}}_M} \end{bmatrix} = 0,
\end{equation}
where $\bf{W}$ is a matrix composed of vectors ${\bf{v}}_0$, ${\bf{w}}_2$, ${\bf{w}}_3$, $\cdots$, ${\bf{w}}_M$, i.e., 
\begin{equation}\label{Eq:ConstrainedOpt_W}
{\bf{W}} = \begin{bmatrix}
{\bf{v}}_0  & {\bf{w}}_2  & {\bf{w}}_3 & \cdots  & {\bf{w}}_M
\end{bmatrix},
\end{equation}
where
\begin{equation}\label{Eq:ConstrainedOpt_Wi}
\begin{split}
{{\bf{w}}_i} &= {\bf{K}}{\left( {\xi \cdot {{\bf{v}}_0}{{\bf{v}}_0}^H - {{\bf{D}}_{{f_0}}}} \right)^{ - 1}}{\bf{K}}{{\bf{v}}_i}\\
 &= \left[ {p^2}\left( {\xi \cdot {{\bf{v}}_0}{{\bf{v}}_0}^H - {{\bf{D}}_{{f_0}}}} \right) - 2p{\bf{B}} \right.\\
 &\left.\quad+ {\bf{B}}{{\left( {\xi \cdot {{\bf{v}}_0}{{\bf{v}}_0}^H - {{\bf{D}}_{{f_0}}}} \right)}^{ - 1}}{\bf{B}} \right]{{\bf{v}}_i},\\
&\qquad\qquad\qquad\qquad\qquad\; i = 2,3, \cdots ,M,
\end{split}
\end{equation}
$\{ {{{\bf{v}}_0},{{\bf{v}}_2},{{\bf{v}}_3}, \cdots ,{{\bf{v}}_M}} \}$ is a basis of the vector space $\mathbb{C}^M$, and 
\begin{equation*}
\text{span}\{ {{{\bf{v}}_0}} \} \bot \text{span}\{ {{{\bf{v}}_2},{{\bf{v}}_3}, \cdots ,{{\bf{v}}_M}} \}.
\end{equation*}
\end{theorem}
\begin{proof}
\quad \emph{Proof:}
See Appendix \ref{proof:theoConstrainedOpt_p}.
\end{proof}

Theorem \ref{theoConstrainedOpt_p} formulates a $2\left( {M - 1} \right)$-degree polynomial equation for $p$, which can be numerically solved. One of the solutions gives the global maximum of directivity. This result enables the transformation of an implicit equation containing an unknown parameter into an explicit equation that can be easily solved numerically.

Once $p$ is determined, the matrix $\bf{K}$ and the desired excitation vector $\bf{a}$ can be sequentially calculated using (\ref{Eq:ConstrainedOpt_a_K}) and (\ref{Eq:ConstrainedOpt_a}). And the constrained optimum directivity in theory is given by taking the result of (\ref{Eq:ConstrainedOpt_a}) into (\ref{Eq:EEP_D}).

The complete procedure for calculating the robust superdirective beamforming vector, utilizing Theorem \ref{theoConstrainedOpt_a} and Theorem \ref{theoConstrainedOpt_p}, is summarized in Algorithm \ref{Alg:ConstrainedOptSol}. The pivotal step involves identifying the orthogonal complement of the subspace spanned by ${\bf{v}}_0$,
\begin{algorithm}[h]
\caption{Orthogonal Complement-based Robust Superdirective Beamforming}
\begin{algorithmic}[1]\label{Alg:ConstrainedOptSol}
\STATE{Find $\left( {M - 1} \right)$ vectors from the vector space $\mathbb{C}^M$ to constitute a basis of the orthogonal complement of ${\bf{v}}_0$.}

\STATE{Compute vectors ${{\bf{w}}_2}$, ${{\bf{w}}_3}$, $\cdots$, ${{\bf{w}}_M}$ using (\ref{Eq:ConstrainedOpt_Wi}), where their expressions are parameterized by $p$. According to (\ref{Eq:ConstrainedOpt_W}), the matrix ${\bf{W}}$ is formed.}

\STATE{Solve the equation of the parameter $p$ in (\ref{Eq:ConstrainedOpt_detW}), obtaining $2\left( {M - 1} \right)$ roots.}

\STATE{Plug the $2\left( {M - 1} \right)$ roots into (\ref{Eq:ConstrainedOpt_a_K}), (\ref{Eq:ConstrainedOpt_a}) and (\ref{Eq:EEP_D}) sequentially to calculate $\bf{K}$, $\bf{a}$ and $D$.}

\STATE{Choose the $p$ value that maximizes $D$ and it corresponds to the optimum excitation ${\bf{a}}$.}

\end{algorithmic}
\end{algorithm}

Notably, Algorithm \ref{Alg:ConstrainedOptSol} can also be transferred to determine the excitation for maximum signal-to-noise ratio while imposing a Q-factor constraint \cite{lo1966optimization}. Additionally, no restrictions are imposed on antenna arrays during the previous derivations, and therefore the algorithm applies to arbitrary arrays.

\section{Simulation Results}\label{sec:SimulationResults}
\begin{figure}[b]
  \centering
  \includegraphics[width=2.5in]{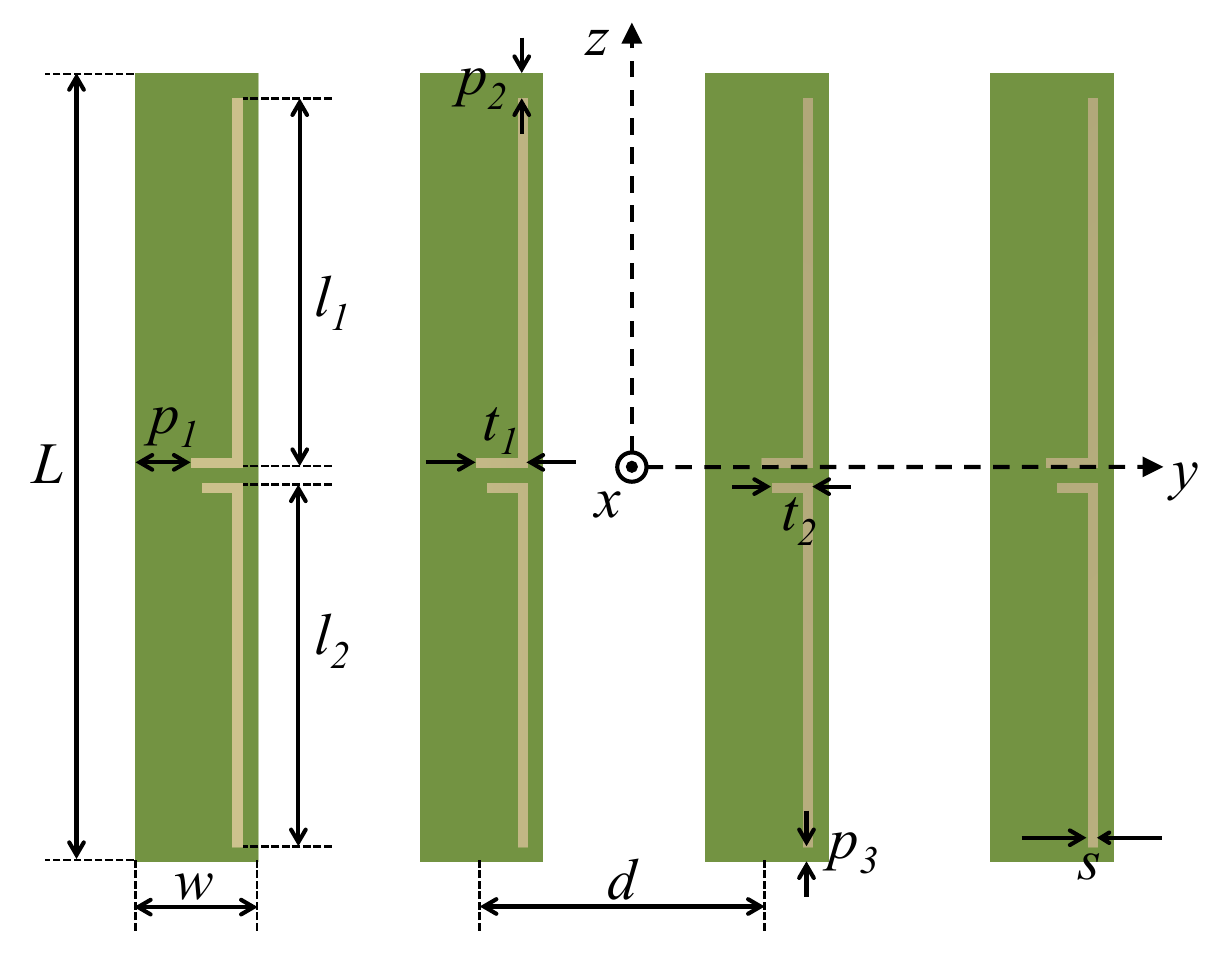}\\
  \caption{The topology of the printed dipole array operating at 1.6 GHz.} \label{fig:SchematicViewOfDipoleArray}
\end{figure}

To validate the proposed EEPB method and the EEP-OCRB method, we have carried out theoretical computations and full-wave simulations. An evenly spaced linear array composed of 4 identical printed dipole antennas operating at 1.6 GHz is designed as shown in Fig. \ref{fig:SchematicViewOfDipoleArray}. 
The direction ${{\bf{u}}_0}$ corresponds to the zenith angle $\theta  = {90^ \circ }$ and the azimuth angle $\phi  = {90^ \circ }$ or $\phi  = {270^ \circ }$, i.e., the endfire direction. The dipole antenna made of annealed copper ($\sigma = 5.8 \times {10^7}$ $S/m$) is printed on a 0.8 $mm$ thick FR-4 substrate (${\epsilon _r} = 4.3$, ${\mu _r} = 1.0$ and $\tan \delta  = 0.025$). The structural parameters are listed in Table \ref{tab:ParametersOfDipole} and their meanings are shown in Fig. \ref{fig:SchematicViewOfDipoleArray}.
The spacing $d$ between the adjacent elements is adjustable. In the following simulations, the dipole antenna array shown in Fig. \ref{fig:SchematicViewOfDipoleArray} is employed by default unless otherwise specified.
\begin{table}[b]
\caption{The Structural Parameters of a Dipole Element\label{tab:ParametersOfDipole}}
\centering
\begin{tabular}{|c|c|}
\hline
\hline
\textbf{Parameter} & \textbf{Value}\\
\hline
Dipole arm 1 & ${l_1} = 36.50$ $mm$, ${t_1} = 5.10$ $mm$\\
\hline
Dipole arm 2 & ${l_2} = 36.00$ $mm$, ${t_2} = 4.06$ $mm$\\
\hline
Copper wires & $s = 1.00$ $mm$\\
\hline
Arms' position & ${p_1} = 5.56$ $mm$, ${p_2} = 2.50$ $mm$, ${p_3} = 1.46$ $mm$\\
\hline
Substrate & $L = 78.00$ $mm$, $w = 12.20$ $mm$\\
\hline
\hline
\end{tabular}
\end{table}

\begin{figure}[t]
  \centering
  \includegraphics[width=3.2in]{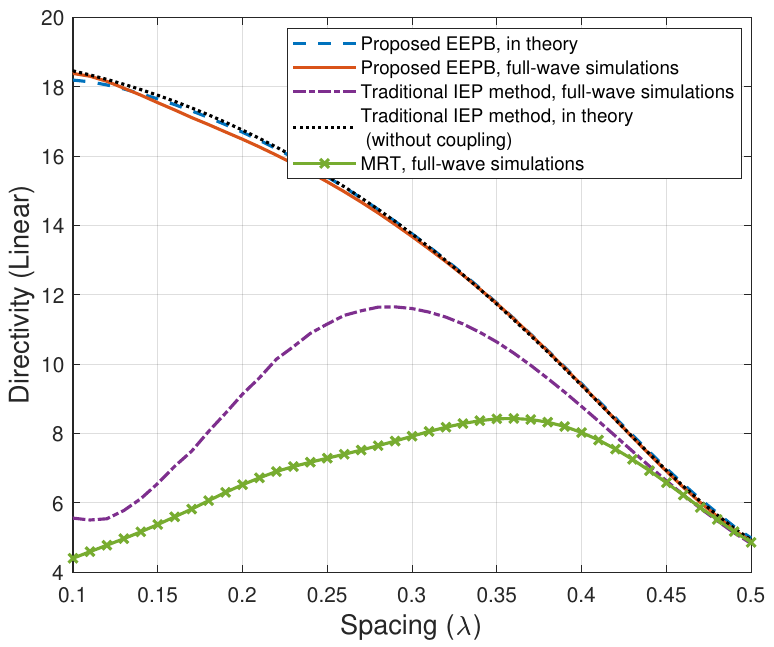}\\
  \caption{The directivity of a 4-dipole antenna array against antenna spacing.} \label{fig:M=4DirvsSpacing}
\end{figure}

We first compare the performance of the proposed EEPB method with two other beamforming methods: the traditional IEP method and the conventional MRT. The spacing $d$ between the elements ranges from $0.10\lambda$ (since the printed dipole antenna has a non-zero width, the spacing cannot be too small) to $0.50\lambda$ at intervals of $0.01\lambda$. The simulation results are illustrated in Fig. \ref{fig:M=4DirvsSpacing}. The results of the EEPB method in theory are calculated by (\ref{Eq:EEP_D0}) and (\ref{Eq:EEP_a0}). The directivity curve of the EEPB method in full-wave simulations is obtained when the excitation vector from (\ref{Eq:EEP_a0}) is applied to the array. The theoretical directivity of the IEP method without coupling is calculated in an imaginary array where no pattern distortion occurs, simply obtained from (\ref{Eq:IEP_pattern}) and (\ref{Eq:D}).

From Fig. \ref{fig:M=4DirvsSpacing}, we can find that all the curves nearly overlap around $0.50\lambda$, while the directivity of the EEPB method becomes higher than both conventional MRT and the traditional IEP method at small spacing, and the gaps continue to increase as the spacing decreases. The curve of the EEPB method plotted via full-wave simulations is in good agreement with the theoretical one, both of which increase as the element spacing decreases. Using the EEPB method, the theoretical directivity factor reaches 18.24 and the directivity factor obtained from full-wave simulations reaches 18.49 when the spacing is $0.10\lambda$, which is indeed ``superdirective". With the spacing narrowing, the directivity of the IEP method and MRT increases at first and then falls down. Moreover, the virtual curve of the IEP method without coupling agrees well with the EEPB method curves and the directivity factors on it are higher than full-wave simulation results of the IEP method. It confirms that the directivity deterioration at small spacing is caused by coupling and the EEPB method has fixed it.

\begin{figure}[t]
  \centering
  \includegraphics[width=3.2in]{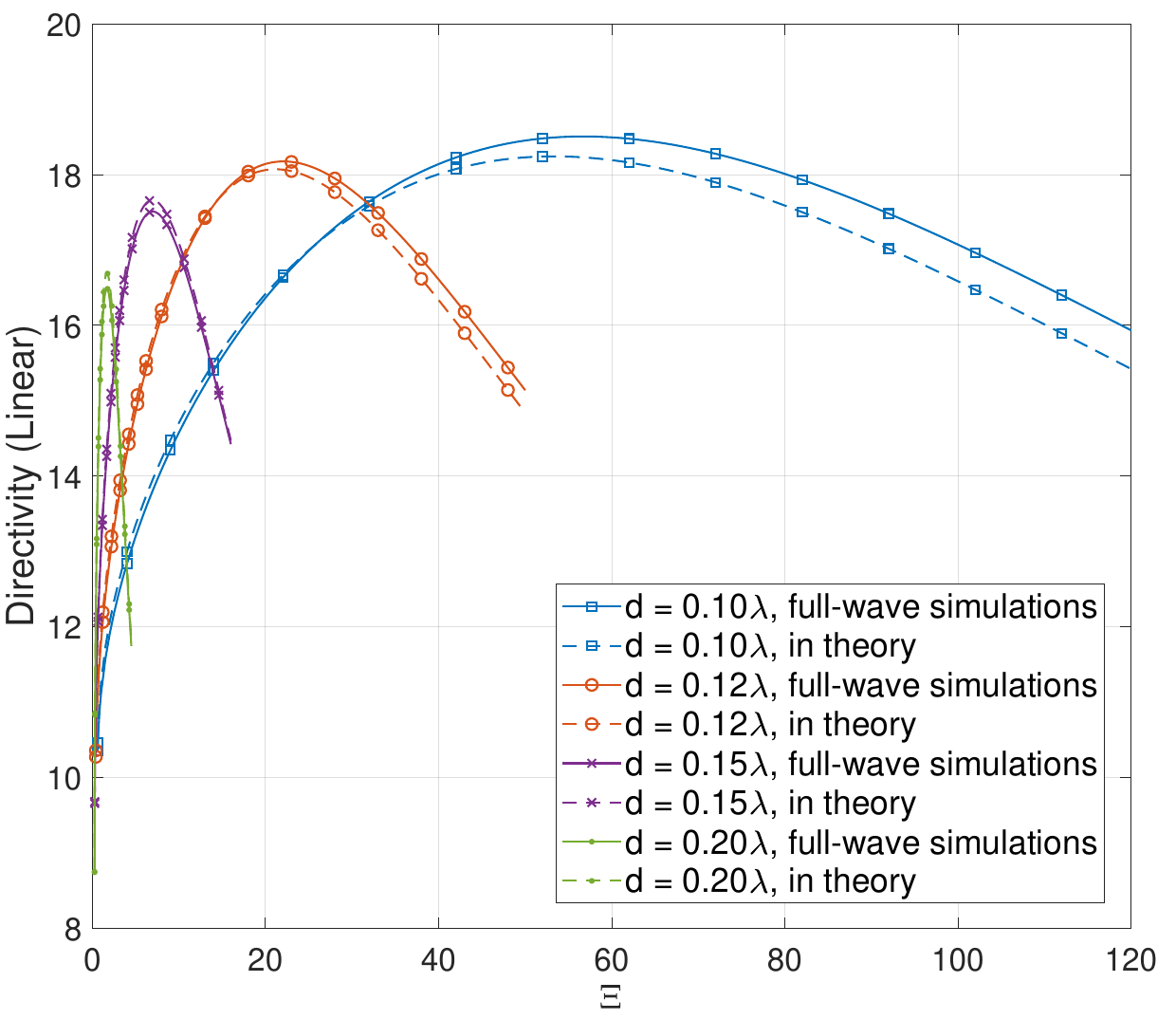}\\
  \caption{The constrained optimum directivity calculated through EEP-OCRB algorithm against the normalized variance $\Xi$, obtained from theoretical computations and full-wave simulations.} \label{fig:M=4DirvsNormalizedVariance_Num}
\end{figure}

Next in Fig. \ref{fig:M=4DirvsNormalizedVariance_Num}, we present the directivity curves against the normalized variance $\Xi$, using the EEP-OCRB method. We show the maximum directivity factors at different spacings under a constraint on the normalized variance $\Xi$, obtained from theoretical computations and full-wave simulations.
For a given value of the constraint $\Xi$, we first calculate the constrained optimum excitation vector ${\bf{a}}$ through Algorithm \ref{Alg:ConstrainedOptSol}. Then we plug it into (\ref{Eq:EEP_D}) in theoretical computation and stimulate the antenna array with it in full-wave simulations to obtain the corresponding directivity factor. For an array of fixed spacing, the constrained optimum directivity first rises then declines as the normalized variance increases. The left starting point of each curve corresponds to the situation where $\Xi$ achieves its minimum value of $\frac{1}{M}$, as described by (\ref{Eq:NormalizedVariancerange}) and (\ref{Eq:ConditionOfNormalizedVariancemin}), exhibiting the least array pattern fluctuations yet a very low directivity factor. Initially, the maximum directivity climbs with the increasing of the normalized variance, sacrificing robustness for high directivity. The critical point, marking the end of the rising segment of the curve, corresponds to the EEPB method, where the upper limit of directivity is achieved. After that point, increasing the normalized variance only results in decreasing directivity. When the spacing decreases, the critical point on the corresponding curve indicates an increase in the directivity and the normalized variance, i.e., better superdirectivity yet poorer robustness under the EEPB method. Robust superdirective beamforming is realized by seeking a suitable trade-off on the rising segment of the curve before the critical point.

We now verify the effectiveness of the EEP-OCRB method and the antenna spacing is set to $d=0.10\lambda$. Fig. \ref{fig:M=4,d=0.10,RobustnessEffect} compares the directivity histograms of the unconstrained EEPB method and the EEP-OCRB method.  
\begin{figure}[t]
  \centering
  \includegraphics[width=3.2in]{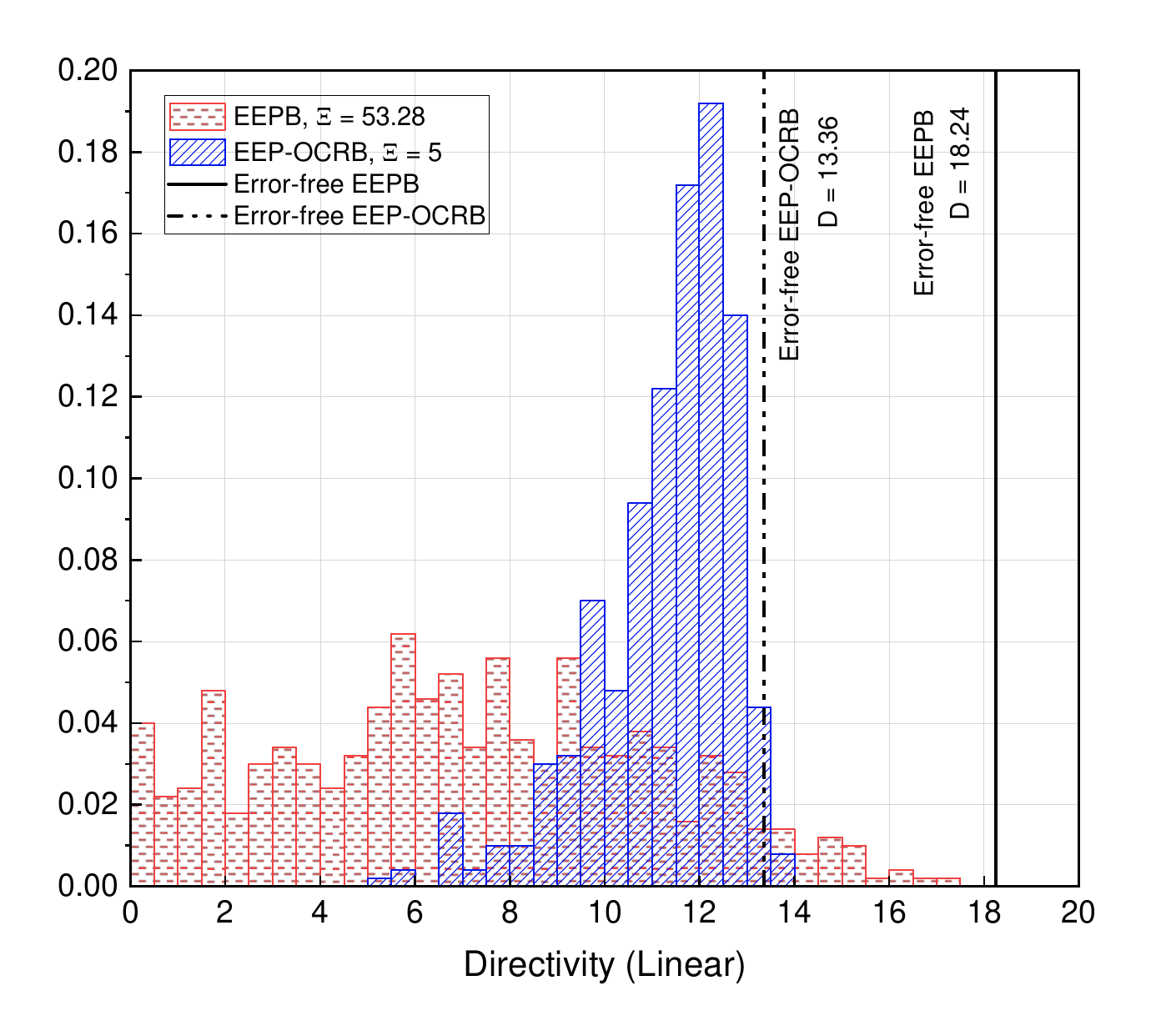}\\
  \caption{The directivity histograms of the unconstrained EEPB method and the EEP-OCRB method when the spacing is $0.10\lambda$, along with the error-free directivity factors. The standard deviation of relative amplitude error is $\sqrt{\overline {{\alpha ^2}}} = 0.05$, the standard deviation of phase error is $\sqrt{\overline {{\delta ^2}}} = {5^ \circ }$, and the number of samples is $N=500$.} \label{fig:M=4,d=0.10,RobustnessEffect}
\end{figure}
The former produces a $\Xi$ value of $53.28$, while the latter is constrained to a $\Xi$ value of $5$. The Monte Carlo simulation settings, including the values of $\sqrt{\overline {{\alpha ^2}}}$, $\sqrt{\overline {{\delta ^2}}}$ and $N$, are identical to those used in Section \ref{sec:SensitivityAnalysis}. The directivity factors generated with error-free excitation vectors are also marked. It can be observed that the EEP-OCRB method produces a significantly more concentrated distribution of directivity factors with the same degree of excitation errors. Adding the constraint of the normalized variance $\Xi$ makes $\mathcal{H}$ reduce from $139.90$ to $6.27$.

\begin{figure}[!t]
\centering
\subfloat[]{\includegraphics[width=2.7in]{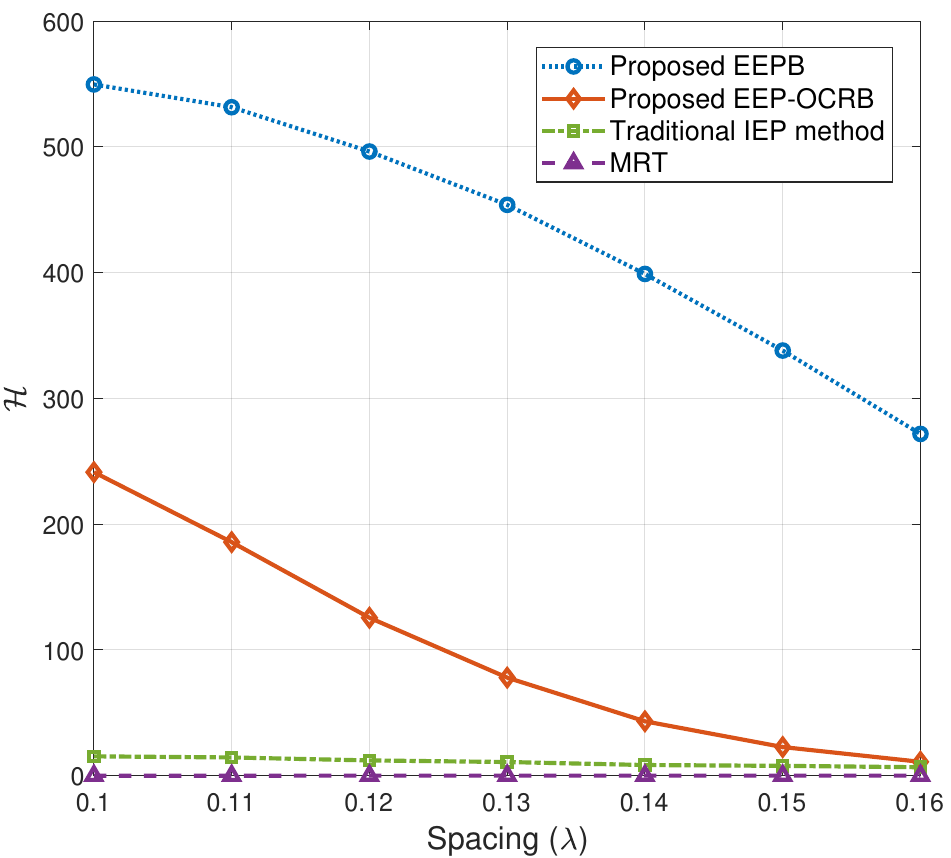}%
\label{fig:H vs. Spacing}}
\\
\subfloat[]{\includegraphics[width=2.7in]{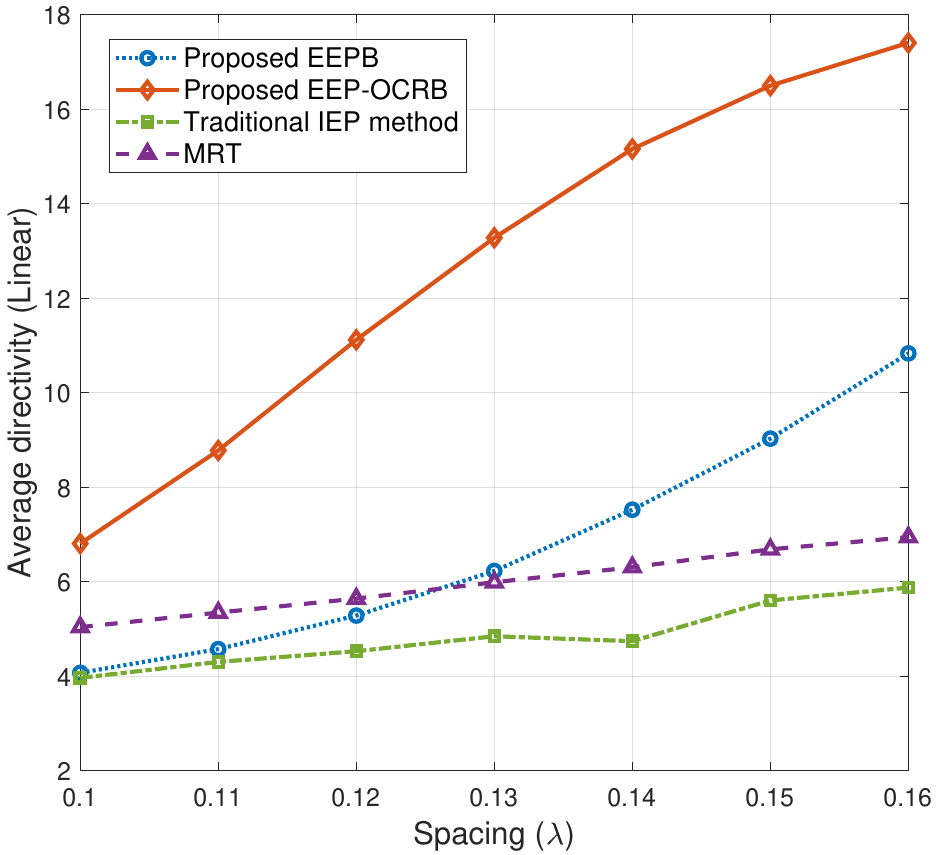}%
\label{fig:Average directivity vs. Spacing}}
\caption{The performances of the unconstrained EEPB method, the EEP-OCRB method, the traditional IEP method and MRT in a five-dipole antenna array. (a) $\mathcal{H}$ against spacing. (b) The average directivity against spacing.}
\label{fig:EEP vs. Robust}
\end{figure}

The validity of the EEP-OCRB method at different antenna spacings is verified in a five-dipole antenna array. The element structure and alignment are the same as those in Fig. \ref{fig:SchematicViewOfDipoleArray}, and the only difference is the number of elements. Monte Carlo simulations are conducted for the unconstrained EEPB method, the EEP-OCRB method, the traditional IEP method, and MRT respectively using the following settings: $\sqrt{\overline {{\alpha ^2}}} = 0.05$, $\sqrt{\overline {{\delta ^2}}} = {5^ \circ }$, and $N=100000$. In Fig. \ref{fig:EEP vs. Robust}, we compare $\mathcal{H}$ which is defined in (\ref{Eq:H}) and the average directivity among these beamforming methods. 
In the EEP-OCRB method, we impose a constraint to decrease the normalized variance $\Xi$ to $10\%$ of that in the unconstrained EEPB method. The EEP-OCRB method we propose exhibit significant reduction in $\mathcal{H}$ compared to the unconstrained EEPB method, resulting in smaller directivity fluctuations with the same extent of excitation variations. In fact, with the increasing of the antenna spacing, $\mathcal{H}$ of the robust methods can be effectively brought close to the level of MRT, which is well-known for its robustness. Furthermore, thanks to the robustness, the methods also show a significant increase in average directivity compared to other beamforming approaches.

\section{Experimental Results}\label{sec:ExperimentalResults}
To validate the effectiveness of the EEPB method and the EEP-OCRB method, we have designed a prototype of superdirective antenna array working at 1.6 GHz. This array is composed of the printed circuit board (PCB) dipole antennas that have the same structures as those in the simulations.

\begin{figure}[h]
  \centering
  \includegraphics[width=3.2in]{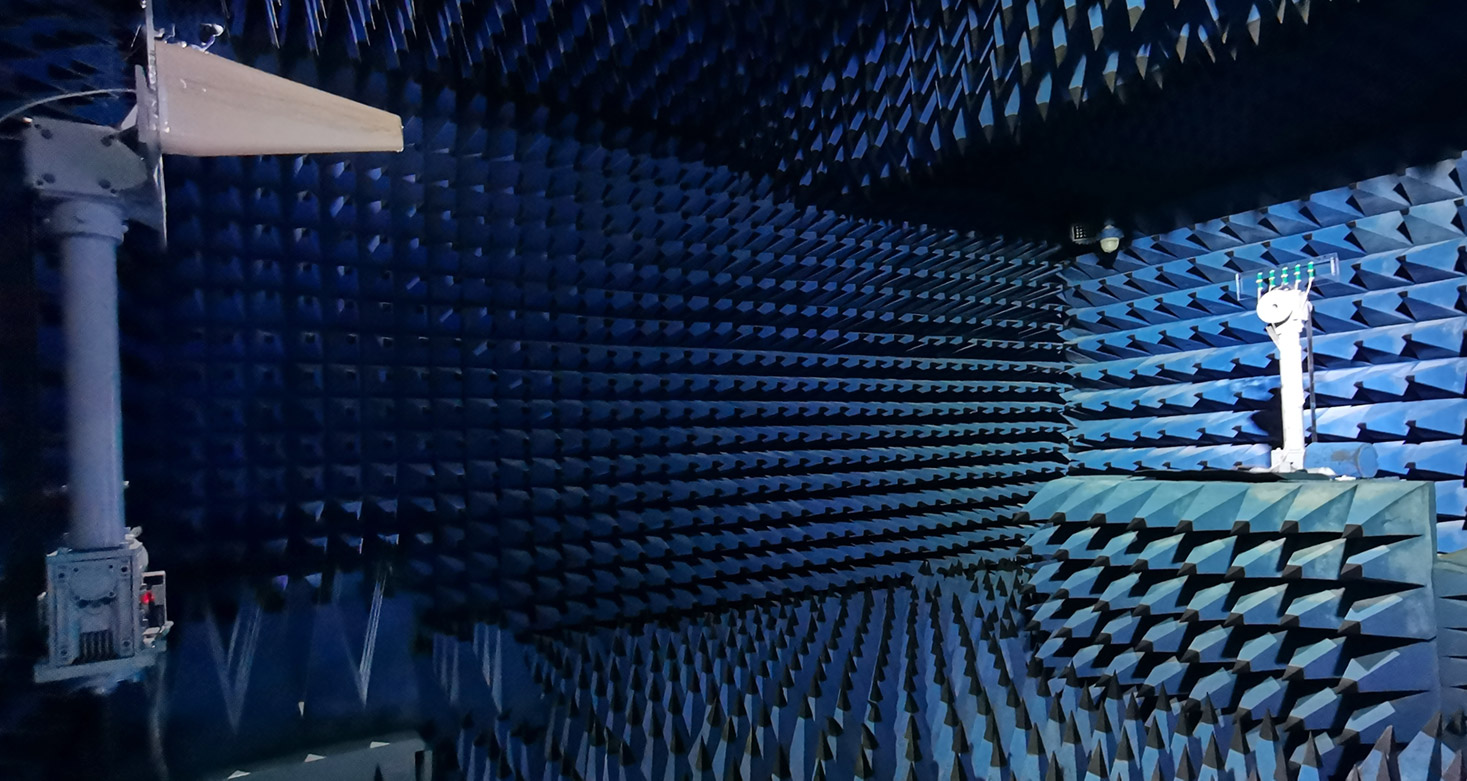}\\
  \caption{The microwave anechoic chamber size is 4 m $\times$ 6 m $\times$ 4 m, and the array-RX antenna distance is 4.8 m.} \label{fig:MicrowaveAnechoicAmber}
\end{figure}

\begin{figure}[h]
  \centering
  \includegraphics[width=3.2in]{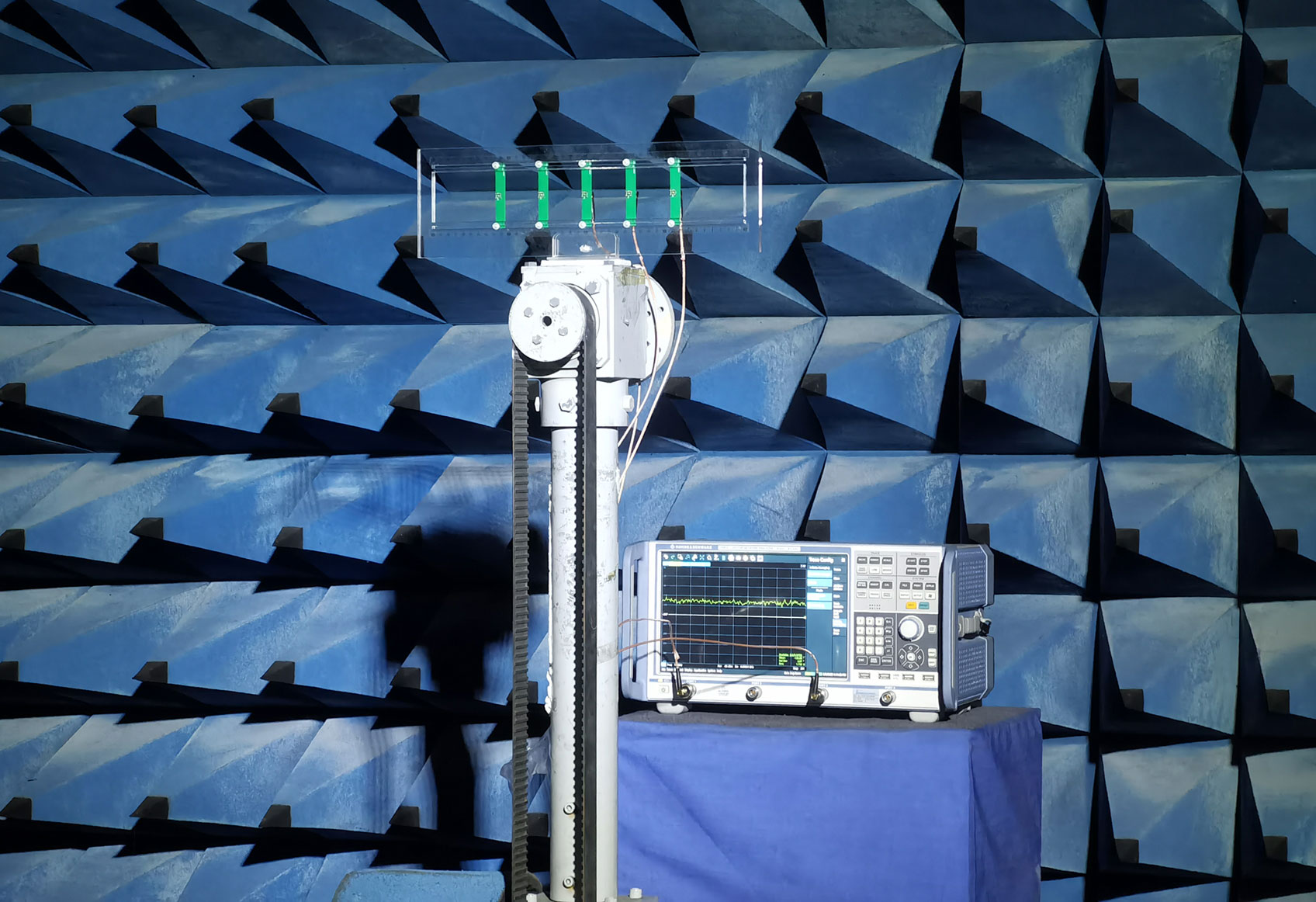}\\
  \caption{The measurement of the initial phase differences between the channels of the beamforming control board.} \label{fig:MeasureInitialPhaseDiff}
\end{figure}

We carry out the experiments in a microwave anechoic chamber, as shown in Fig. \ref{fig:MicrowaveAnechoicAmber}. 
The antenna array is installed on a rotating platform. An 8-channel beamforming control board, which was used in \cite{han2023superdirective}, is utilized here. It enables each antenna elements to be excited with software-controlled excitation coefficients, quantizing the amplitude with 7 bits and the phase with 8 bits. The initial phase differences between the channels of the beamforming control board are measured and calibrated in the software. Fig. \ref{fig:MeasureInitialPhaseDiff} shows the measurement process. 
In the experiments, we use a VNA to measure the farfield radiation pattern.

By stimulating the elements in the array one by one with uniform excitation, we obtain the EEP of each element. The excitaion coefficients are computed through the measured EEPs. Then we excite the array with them through the beamforming control board and measure the array power pattern. Since the rotating platform can only rotate on the plane, the EEPs and the array patterns are all measured on a certain plane. Unavoidably, the absence of the radiation information on the other planes makes the effects of beamforming compromised. Nevertheless, this can be solved with a three-dimensional rotating platform. To compare the directivity of the planar patterns, we define the planar directivity factor as:
\begin{equation}\label{Eq:D_p}
{D_p} \triangleq \frac{{\mathop {\max }\limits_\phi  {{\left| {{\bf{F}}\left( {{\theta _0},\phi } \right)} \right|}^2}}}{{\frac{1}{{2\pi }}\int_0^{2\pi } {{{\left| {{\bf{F}}\left( {{\theta _0},\phi } \right)} \right|}^2}d\phi } }},
\end{equation}
where $\theta = \theta_0$ is the selected plane.

\begin{figure}[h]
  \centering
  \includegraphics[width=2.5in]{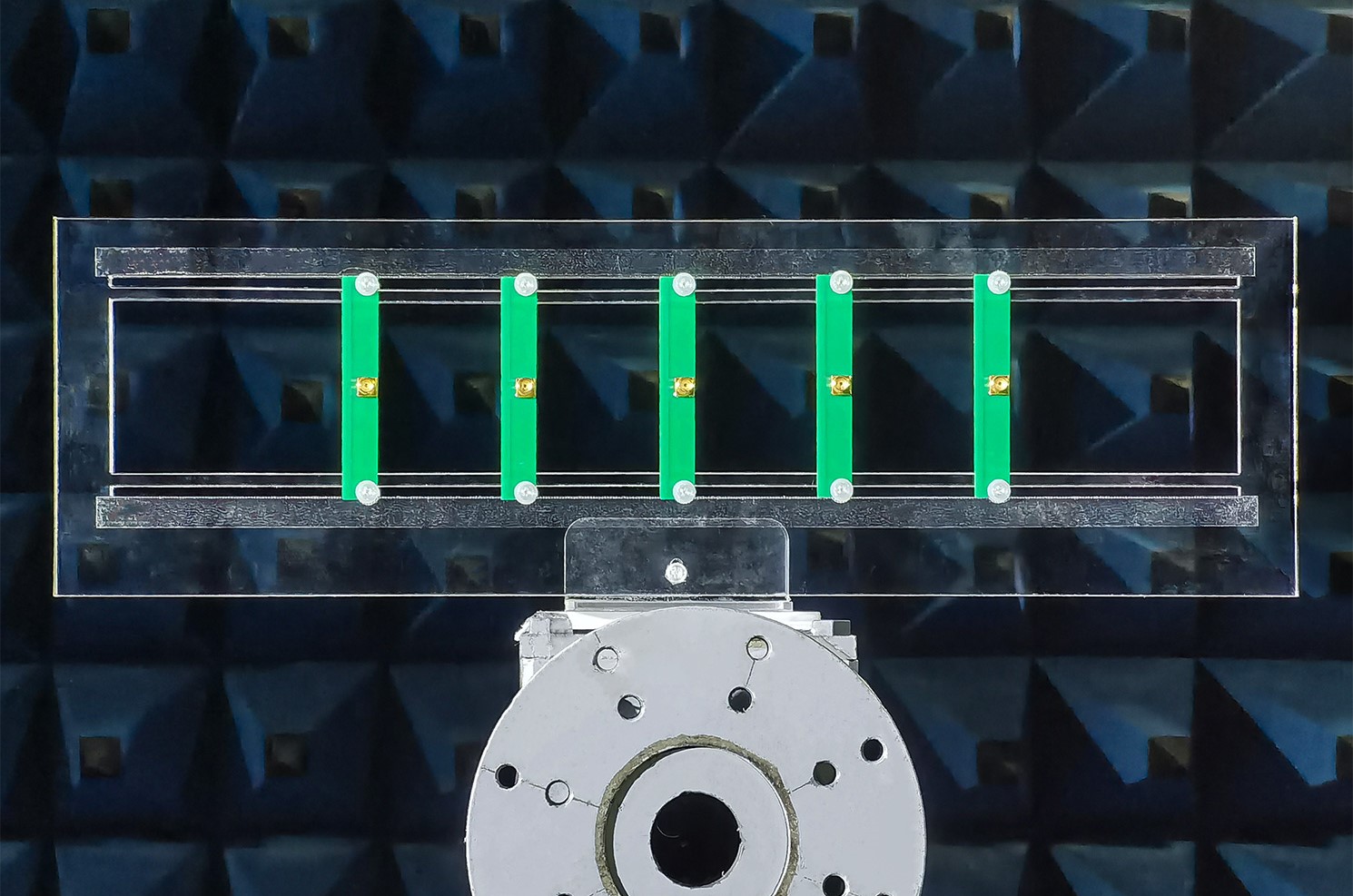}\\
  \caption{The antenna array under test.} \label{fig:AUT}
\end{figure}

In the experiments, the number of antennas is $M = 5$, the spacing between the elements is set to $d = 0.3\lambda$, and the testing plane is selected as $\theta_0 = 90^{\circ}$. The antenna array under test is presented in Fig. \ref{fig:AUT}. 
We set the direction of maximum radiation as the endfire direction $\phi_0 = 270^\circ$.

\begin{figure*}[t]
\centering
\subfloat[]{\includegraphics[width=2.3in]{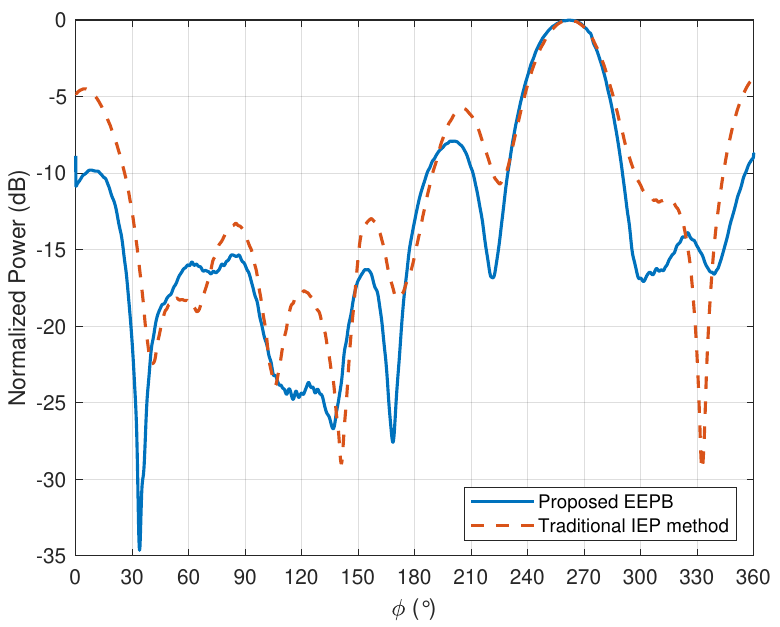}%
\label{fig:ExpValEEP_EEPvsIEP}}
\hfil
\subfloat[]{\includegraphics[width=2.3in]{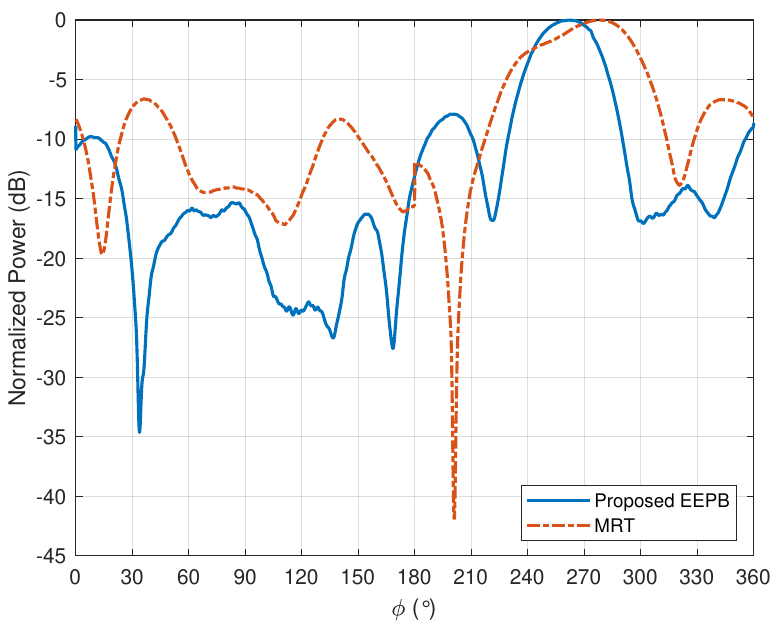}%
\label{fig:ExpValEEP_EEPvsMRT}}
\hfil
\subfloat[]{\includegraphics[width=2.3in]{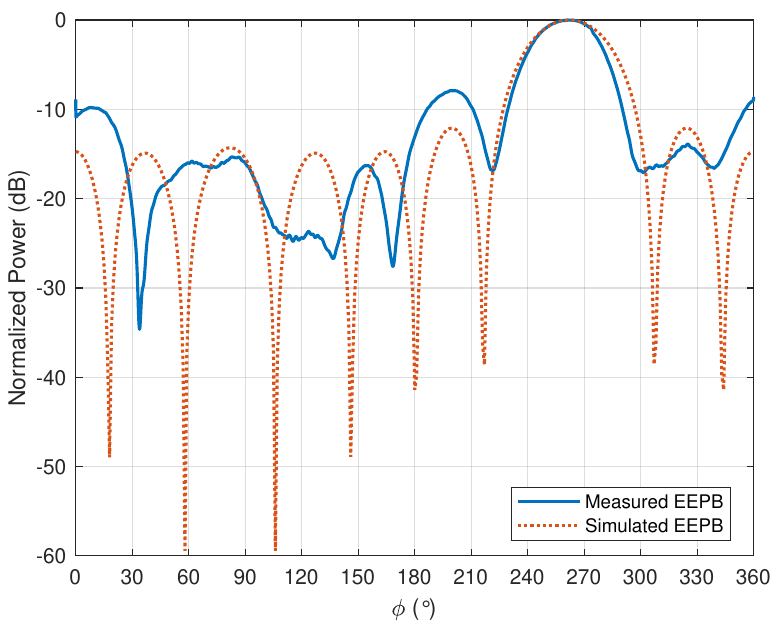}%
\label{fig:ExpValEEP_MeasuredvsSim}}
\caption{The measured normalized power pattern of the propsed EEPB method on the testing plane $\theta = 90^\circ$ compared with (a) the traditional IEP method, (b) the MRT method and (c) the simulated EEPB method.}
\label{fig:ExpValEEP}
\end{figure*}

We first validate the superdirectivity of the proposed EEPB method. Three kinds of beamforming methods are employed in the experiments: the EEPB method, the traditional IEP method and MRT. Besides, we also excite the same antenna array using the EEPB method in the full-wave simulation. The comparions of their normalized power patterns in the testing plane are presented in Fig. \ref{fig:ExpValEEP}. 
\begin{figure*}[t]
\centering
\subfloat[]{\includegraphics[width=2.3in]{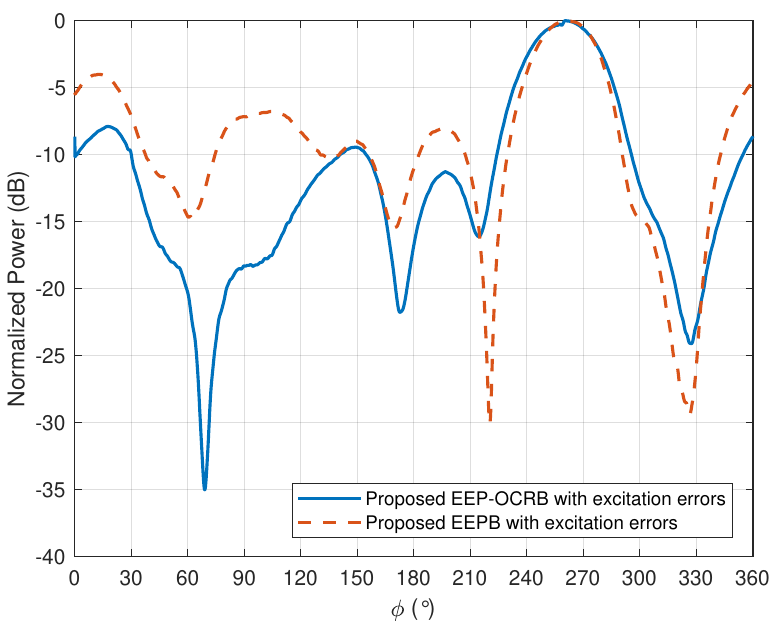}%
\label{fig:ExpValRobust_RobustvsEEP}}
\hfil
\subfloat[]{\includegraphics[width=2.3in]{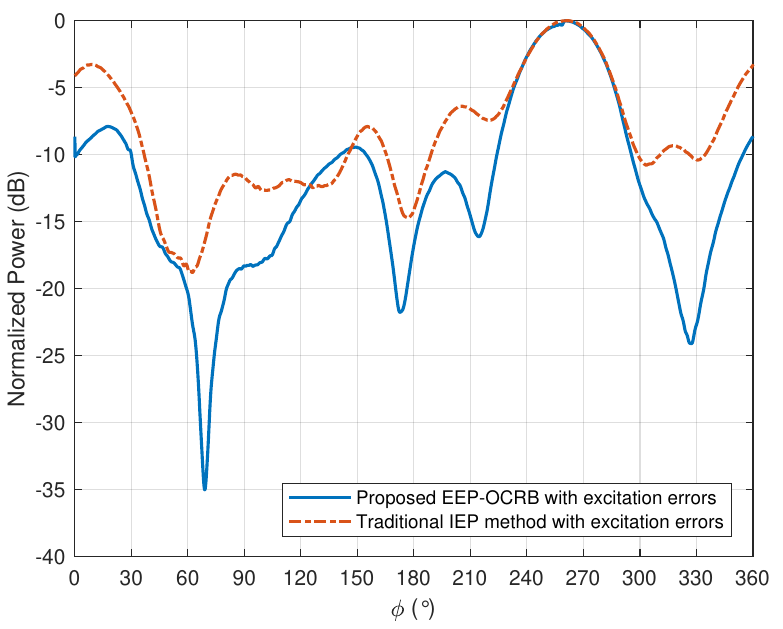}%
\label{fig:ExpValRobust_RobustvsIEP}}
\hfil
\subfloat[]{\includegraphics[width=2.3in]{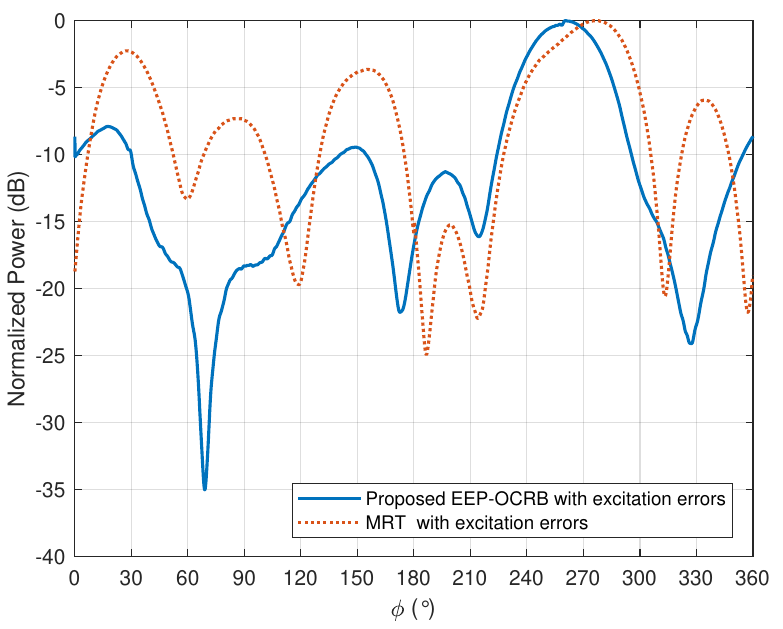}%
\label{fig:ExpValRobust_RobustvsMRT}}
\caption{The measured normalized power pattern of the EEP-OCRB method with excitation errors on the testing plane $\theta = 90^\circ$ compared with (a) the unconstrained EEPB method with excitation errors, (b) the traditional IEP method with excitation errors and (c) the MRT method with excitation errors.}
\label{fig:ExpValRobust}
\end{figure*}
Compared to the traditional method, our proposed method has nearly the same beamwidth in the main lobe, yet lower sidelobes, resulting in higher directivity. On the other hand, the MRT method exhibits a much wider main lobe and higher sidelobes. Furthermore, the measured pattern of the EEPB method closely aligns with the simulated pattern, affirming the validity of our aforementioned theory. Some discrepencies may be attributed to the coaxial feed lines and the SubMiniature version A (SMA) connectors connected to the antennas on one side, which affect the radiation characteristics of the antenna array. Two criteria for these patterns---the planar directivity factor $D_p$ and the half power beamwidth (HPBW)---are listed in Table \ref{tab:CriteriaOfExpValEEP}. 
Clearly, the proposed EEPB method outperforms the others with its high $D_p$ and narrow beamwidth.

We then verify the robustness of the EEP-OCRB method. We add identical errors to the excitation coefficients in the EEP-OCRB method, the unconstrained EEPB method, the traditional IEP method and MRT, respectively. These excitation errors are sampled from the distribution described in (\ref{Eq:NormalDistribution}), where $\sqrt{\overline {{\alpha ^2}}} = 0.10$ and $\sqrt{\overline {{\delta ^2}}} = {20^ \circ }$. Fig. \ref{fig:ExpValRobust} shows the normalized power patterns of the above beamforming methods with errors.
Despite having similar main lobe beamwidths, the EEP-OCRB method with excitation errors maintains lower sidelobes compared to the other methods with the same errors. The criteria for their patterns are listed in Table \ref{tab:CriteriaOfExpValRobust}. It can be observed that, with the same excitation errors, the EEP-OCRB method maintains the highest $D_p$, while its beamwidth is only slightly wider. It is confirmed that our proposed EEP-OCRB method is less susceptible to excitation errors and exhibits robust superdirectivity.
\begin{table}[!t]
\caption{The Criteria for Measured Power Patterns under Different Beamforming Methods\label{tab:CriteriaOfExpValEEP}}
\centering
\begin{tabular}{|c|c|c|c|c|}
\hline
\hline
\textbf{Criteria} & \textbf{\makecell[c]{Proposed \\ EEPB method}} & \textbf{\makecell[c]{Traditional \\ IEP method}} & \textbf{MRT} & \textbf{\makecell[c]{Simulated \\ EEPB method}}\\
\hline
$D_p$ & $7.2467$ & $5.6789$ & $4.5399$ & $7.2645$\\
\hline
HPBW & $37.80^\circ$ & $37.22^\circ$ & $61.97^\circ$ & $44^\circ$\\
\hline
\hline
\end{tabular}
\end{table}
\begin{table}[!t]
\caption{The Criteria for Measured Power Patterns under Different Beamforming Methods with Excitation Errors\label{tab:CriteriaOfExpValRobust}}
\centering
\begin{tabular}{|c|c|c|c|c|}
\hline
\hline
\textbf{Criteria} & \textbf{\makecell[c]{Proposed EEP- \\ OCRB method}} & \textbf{\makecell[c]{Proposed \\ EEPB method}} & \textbf{\makecell[c]{Traditional \\ IEP method}} & \textbf{MRT}\\
\hline
$D_p$ & $6.3582$ & $4.8285$ & $4.3211$ & $3.8970$\\
\hline
HPBW & $42.32^\circ$ & $37.41^\circ$ & $42.51^\circ$ & $48.74^\circ$\\
\hline
\hline
\end{tabular}
\end{table}

\section{Conclusions}\label{sec:conclusions}
In this paper, we proposed a superdirective beamforming method based on the EEP, which contains the mutual-coupling information. In this EEPB method, the beamforming vector for strongly coupled arrays is derived by incorporating the EEP into the model of array pattern. This enables a more accurate description of the radiation field compared to the conventional pattern multiplication rule. We then proposed a robust superdirective beamforming method on the basis of evaluating the sensitivity and the EEP, searching for a trade-off between the directivity and the sensitivity. An EEP-OCRB algorithm was proposed for calculating the robust excitation vector. 
Additionally, we carried out simulations and designed a prototype of superdirective antenna array for real-world experiments. The antennas operate at 1.6 GHz. The simulation results and experimental results both demonstrated that the EEPB method outperforms other beamforming methods and validated the effectiveness of the EEP-OCRB approach.

%%%%%%%%%%%%%%%%%%%%%%%%%%%%%%%%%%%%%%%%%%%%%%%%%%%%%%%%%%%%
\appendix
\subsection{Proof of Theorem \ref{theoRayleighRitz}}\label{proof:theoRayleighRitz}
\begin{proof}
Assume
\begin{equation}\label{Eq:EEP_Lagrg_Const}
{{\bf{a}}^T}{\bf{B}}{{\bf{a}}^*} = C,
\end{equation}
where $C$ is a constant. Then we use the method of Lagrange multipliers and derive a new function:
\begin{equation}\label{Eq:EEP_Lagrg_L}
L = {{\bf{a}}^T}{{\bf{v}}_0}{{\bf{v}}_0}^H{{\bf{a}}^*} - \mu \left( {{{\bf{a}}^T}{\bf{B}}{{\bf{a}}^*} - C} \right),
\end{equation}
where $\mu$ is the Lagrange multiplier we introduced.The first variation of $L$ is
\begin{multline}\label{Eq:EEP_Lagrg_VariationL1}
\delta L = \delta {{\bf{a}}^T}\left( {{{\bf{v}}_0}{{\bf{v}}_0}^H{{\bf{a}}^*} - \mu {\bf{B}}{{\bf{a}}^*}} \right) \\
+ \left( {{{\bf{a}}^T}{{\bf{v}}_0}{{\bf{v}}_0}^H - \mu {{\bf{a}}^T}{\bf{B}}} \right)\delta {{\bf{a}}^*}.
\end{multline}
Since ${\bf{B}}$ is Hermitian, Eq. (\ref{Eq:EEP_Lagrg_VariationL1}) can be rewritten as
\begin{equation}\label{Eq:EEP_Lagrg_VariationL2}
\delta L = 2\Re \left\{ {\delta {{\bf{a}}^T}\left( {{{\bf{v}}_0}{{\bf{v}}_0}^H{{\bf{a}}^*} - \mu {\bf{B}}{{\bf{a}}^*}} \right)} \right\}.
\end{equation}
Hence, Eq. (\ref{Eq:EEP_Lagrg_L}) reaches its maximum value when $\delta L$ is equal to zero, i.e.,
\begin{equation}\label{Eq:EEP_GeneralizedEigenvalue}
{{\bf{v}}_0}{{\bf{v}}_0}^H{{\bf{a}}^*} = \mu {\bf{B}}{{\bf{a}}^*},
\end{equation}
which turns into a generalized eigenvalue problem. Therefore, the solution of the optimization problem, i.e., the maximum directivity factor and the corresponding excitation vector are (\ref{Eq:EEP_D0}) and (\ref{Eq:EEP_a0}) respectively. This concludes the proof.
\end{proof}

\subsection{Beam Coupling Factors}\label{proof:BCF}
\begin{proof}
The electric field of the $i$-th element in the farfield region is written as:
\begin{equation}
{{\bf{E}}_i}\left( {\bf{u}} \right) = {{\bf{f}}_i}\left( {\bf{u}} \right)\frac{{{e^{ - jkr}}}}{r}{V_{g,i}},
\end{equation}
where $r$ is the distance from the origin to the field point. When all the antenna elements are excited, the total electric field is
\begin{equation}
{\bf{E}}\left( {\bf{u}} \right) = \sum\limits_{i = 1}^M {{{\bf{E}}_i}\left( {\bf{u}} \right)}  = \sum\limits_{i = 1}^M {{{\bf{f}}_i}\left( {\bf{u}} \right)\frac{{{e^{ - jkr}}}}{r}{V_{g,i}}}.
\end{equation}
Then the time-averaged radiated power of this M-antenna array can be formulated in terms of the EEPs as:
\begin{equation}
P = \frac{1}{{2\eta }}\int_S {{{\left| {\sum\limits_{i = 1}^M {{{\bf{f}}_i}\left( {\bf{u}} \right){V_{g,i}}} } \right|}^2}d\Omega }.
\end{equation}
Integrating over the unit sphere $\Omega$, one has:
\begin{equation}\label{Eq:PowerRad_Poynting}
P = \frac{{4\pi }}{{2\eta }}{{\bf{V}}_g}^T{{\bf{BV}}_g}^*,
\end{equation}
where
\begin{equation}
{\bf{V}_g} = \begin{bmatrix}
V_{g,1}  & V_{g,2}  & \cdots  & V_{g,M}
\end{bmatrix}^T.
\end{equation}

The power leaving the volume formed by all generators can also be represented in terms of the power waves $\bf{a}$ and $\bf{b}$ as:
\begin{equation}\label{Eq:PowerRad_PowerWaves}
P = \frac{1}{2}\left( {{{\bf{a}}^T}{{\bf{a}}^*} - {{\bf{b}}^T}{{\bf{b}}^*}} \right),
\end{equation}
where
\begin{align}
{\bf{a}} &= \begin{bmatrix}
a_1  & a_2  & \cdots  & a_M
\end{bmatrix}^T\\
{\bf{b}} &= \begin{bmatrix}
b_1  & b_2  & \cdots  & b_M
\end{bmatrix}^T,
\end{align}
and $a_i$ and $b_i$ are the power waves at the $i$-th port. When all ports share the same real-valued reference impedance $Z_0$, the following relationships hold true:
\begin{align}
{\bf{a}} &= \frac{\bf{V}_g}{{2\sqrt {\Re \left\{ {{Z_0}} \right\}} }}\\
{\bf{b}} &= {\bf{Sa}}.
\end{align}
Hence, one obtains:
\begin{equation}\label{Eq:PowerRad_NormS}
P = \frac{1}{{8\Re \left\{ {{Z_0}} \right\}}}{{\bf{V}}_g}^T\left( {{{\mathbf{I}}_M} - {{\bf{S}}^T}{{\bf{S}}^*}} \right){{\bf{V}}_g}^*.
\end{equation}
The property of losslessness implies an equivalence between (\ref{Eq:PowerRad_Poynting}) and (\ref{Eq:PowerRad_NormS}), which consequently leads to (\ref{Eq:B(S)}).
When the reference impedances at each port are different and complex, modified relationships are established:
\begin{align}
{\bf{a}} &= \frac{{{{\bf{V}}_g}}}{{2\sqrt {\Re \left\{ {{{\bf{Z}}_0}} \right\}} }}\\
{\bf{b}} &= {\bf{S}}_{G}\bf{a}.
\end{align}
Thus, in this case (\ref{Eq:PowerRad_PowerWaves}) can be expanded into:
\begin{equation}\label{Eq:PowerRad_GeneralizedS}
P = \frac{1}{8}{{\bf{V}}_g}^T\Re {\left\{ {{{\bf{Z}}_0}} \right\}^{ - \frac{1}{2}}}\left( {{{\mathbf{I}}_M} - {{{{\bf{S}}_G}^T}}{{{{\bf{S}}_G}^*}}} \right)\Re {\left\{ {{{\bf{Z}}_0}} \right\}^{ - \frac{1}{2}}}{{\bf{V}}_g}^*.
\end{equation}
Again, the comparison between (\ref{Eq:PowerRad_Poynting}) and (\ref{Eq:PowerRad_GeneralizedS}) results in (\ref{Eq:B(GeneralizedS)}).

Finally, the power leaving the volume formed by all generators can also be calculated using the circuit theory, expressed in terms of the array impedance matrix $\bf{Z}$ as:
\begin{equation}\label{Eq:PowerRad_CirTheo}
P = \frac{1}{2}{{\bf{I}}^T}\Re \left\{ {\bf{Z}} \right\}{{\bf{I}}^*},
\end{equation}
where 
\begin{equation}
\bf{I} = \begin{bmatrix}
i_1  & i_2  & \cdots  & i_M
\end{bmatrix}^T.
\end{equation}
Since
\begin{equation}
{\bf{I}} = {\left( {{\bf{Z}} + {{\bf{Z}}_0}} \right)^{ - 1}}{{\bf{V}}_g},
\end{equation}
(\ref{Eq:PowerRad_CirTheo}) is expanded into:
\begin{equation}\label{Eq:PowerRad_Z}
P = \frac{1}{2}{{\bf{V}}_g}^T{\left( {{{\left( {{\bf{Z}} + {{\bf{Z}}_0}} \right)}^{ - 1}}} \right)^T}\Re \left\{ {\bf{Z}} \right\}{\left( {{{\left( {{\bf{Z}} + {{\bf{Z}}_0}} \right)}^{ - 1}}} \right)^*}{{\bf{V}}_g}^*.
\end{equation}
Eq. (\ref{Eq:B(Z)}) is obtained by equalizing (\ref{Eq:PowerRad_Poynting}) and (\ref{Eq:PowerRad_Z}).
\end{proof}

\subsection{Proof of Theorem \ref{theoNormalizedVariance}}\label{proof:theoNormalizedVariance}
\begin{proof}
Taking the excitation errors into account, the array pattern function in (\ref{Eq:EEP_pattern}) is rewritten as
\begin{equation}\label{Eq:EEP_PatternWithErrors}
{\bf{F}}\left( {\bf{u}} \right) = \sum\limits_{i = 1}^M {{a_i}\left( {1 + {\alpha _i}} \right){e^{j{\delta _i}}}{{\bf{f}}_i}\left( {\bf{u}} \right)}.
\end{equation}
Then we have
%\begin{equation}
\begin{align}\label{Eq:EEP_PowerPatternWithErrors}
{\left| {{\bf{F}}\left( {\bf{u}} \right)} \right|^2} &= {\left| {\sum\limits_{i = 1}^M {{a_i}\left( {1 + {\alpha _i}} \right){e^{j{\delta _i}}}{{\bf{f}}_i}\left( {\bf{u}} \right)} } \right|^2}\\
&= \sum\limits_{i = 1}^M \sum\limits_{j = 1}^M {a_i}{{a_j}^*}\left( {1 + {\alpha _i}{\alpha _j} + {\alpha _i} + {\alpha _j}} \right) \nonumber \\
&\quad\cdot e^{j ( {{\delta _i} - {\delta _j}} )} {\bf{f}}_i( {\bf{u}})  {{\bf{f}}_j}^H ( {\bf{u}} ).
\end{align}
%\end{equation}
The nominal or no-error array pattern is
\begin{equation}\label{Eq:EEP_PatternWithoutErrors}
{{{\bf{F}}_0}\left( {\bf{u}} \right)} = \sum\limits_{i = 1}^M {{a_i}{{\bf{f}}_i}\left( {\bf{u}} \right)}.
\end{equation}
The no-error power pattern is
\begin{equation}\label{Eq:EEP_PowerPatternWithoutErrors}
\begin{split}
{\left| {{{\bf{F}}_0}\left( {\bf{u}} \right)} \right|^2} &= {\left| {\sum\limits_{i = 1}^M {{a_i}{{\bf{f}}_i}\left( {\bf{u}} \right)} } \right|^2} \\
&= \sum\limits_{i = 1}^M {\sum\limits_{j = 1}^M {{a_i}{{a_j}^*}{{\bf{f}}_i}\left( {\bf{u}} \right){{\bf{f}}_j}^H\left( {\bf{u}} \right)} }.
\end{split}
\end{equation}
Now we calculate the expected values of the field and power patterns. They can be regarded as averages taken over a large number of antenna arrays or over a long time in a single array where the parameters vary with time in a random manner \cite{gilbert1955optimum}. With $\mathbb{E}\left\{ {{\alpha _i}} \right\} = \mathbb{E}\left\{ {{\delta _i}} \right\} = 0$, we have
\begin{equation}\label{Eq:EEP_ExpectedPattern}
\begin{split}
\mathbb{E}\left\{ {{\bf{F}}\left( {\bf{u}} \right)} \right\}  &= \sum\limits_{i = 1}^M {{a_i}\left\{ {1 + \mathbb{E}\left\{ {{\alpha _i}} \right\}} \right\}\mathbb{E}\left\{ {{e^{j{\delta _i}}}} \right\}{{\bf{f}}_i}\left( {\bf{u}} \right)}  \\
&= \sum\limits_{i = 1}^M {{a_i}\left\{ {\mathbb{E}\left\{ {\cos {\delta _i}} \right\} + j\mathbb{E}\left\{ {\sin {\delta _i}} \right\}} \right\}{{\bf{f}}_i}\left( {\bf{u}} \right)},
\end{split}
\end{equation}
where the probability density function of ${\delta _i}$ is
\begin{equation}\label{Eq:delta_i_pdf}
f\left( {{\delta _i}} \right) = \frac{1}{{\sqrt {2\pi \overline {{\delta ^2}} } }}\exp \left( { - \frac{{{\delta _i}^2}}{{2\overline {{\delta ^2}} }}} \right).
\end{equation}
Thus,
\begin{align}
\mathbb{E}\left\{ {\cos {\delta _i}} \right\}  &= \int_{ - \infty }^{ + \infty } {\cos {\delta _i} \cdot f\left( {{\delta _i}} \right)d{\delta _i}}  = {{\mathop{\rm e}\nolimits} ^{ - \frac{{\overline {{\delta ^2}} }}{2}}}\label{Eq:expected_cos_delta_i}\\
\mathbb{E}\left\{ {\sin {\delta _i}} \right\}  &= \int_{ - \infty }^{ + \infty } {\sin {\delta _i} \cdot f\left( {{\delta _i}} \right)d{\delta _i}}  = 0.\label{Eq:expected_sin_delta_i}
\end{align}
By substituting (\ref{Eq:expected_cos_delta_i}) and (\ref{Eq:expected_sin_delta_i}) into (\ref{Eq:EEP_ExpectedPattern}), one has
\begin{equation}\label{Eq:EEP_ExpectedPattern_Final}
\mathbb{E}\left\{ {{\bf{F}}\left( {\bf{u}} \right)} \right\}  = {{\bf{F}}_0}\left( {\bf{u}} \right){{\mathop{\rm e}\nolimits} ^{ - \frac{{\overline {{\delta ^2}} }}{2}}}.
\end{equation}
The expected value of the power pattern is more complicated and shown below:
\begin{equation}\label{Eq:EEP_ExpectedPowerPattern}
\begin{split}
\mathbb{E}\left\{ \left| {\bf{F}\left( {\bf{u}} \right)} \right|^2 \right\}
&= \sum\limits_{i = 1}^M \sum\limits_{j = 1}^M {a_i}a_j^*\left\{ 1 + \mathbb{E}\left\{ \alpha_i\alpha_j \right\} + \mathbb{E}\left\{ \alpha_i \right\} \right.\\
&\quad\left. + \mathbb{E}\left\{ \alpha_j \right\} \right\}\mathbb{E}\left\{ e^{j(\delta_i - \delta_j)} \right\} {\bf{f}}_i\left( {\bf{u}} \right){\bf{f}}_j^H\left( {\bf{u}} \right) \\
&= \sum\limits_{i = 1}^M \sum\limits_{j = 1}^M {a_i}a_j^* \mathbb{E}\left\{ e^{j(\delta_i - \delta_j)} \right\} {\bf{f}}_i\left( {\bf{u}} \right){\bf{f}}_j^H\left( {\bf{u}} \right) \\
&\quad+ \sum\limits_{i = 1}^M \overline {\alpha_i^2} {a_i}a_i^* {\bf{f}}_i\left( {\bf{u}} \right){\bf{f}}_i^H\left( {\bf{u}} \right) \\
&= \sum\limits_{i = 1}^M \sum\limits_{j = 1}^M {a_i}a_j^*{\bf{f}}_i\left( {\bf{u}} \right){\bf{f}}_j^H\left( {\bf{u}} \right)\left\{ \mathbb{E}\left\{ \cos y \right\} \right.\\
&\quad\left.+ j\mathbb{E}\left\{ \sin y \right\} \right\} + \overline {\alpha^2} \sum\limits_{i = 1}^M \left| {a_i} \right|^2 \left| {\bf{f}}_i\left( {\bf{u}} \right) \right|^2,
\end{split}
\end{equation}
where $y = {\delta _i} - {\delta _j}$ is the difference between two identically normal distributed random variables. The distribution of $y$ is
\begin{equation}\label{Eq:y_pdf}
f\left( y \right) = \frac{1}{{\sqrt {4\pi \overline {{\delta ^2}} } }}\exp \left( { - \frac{{{y^2}}}{{4\overline {{\delta ^2}} }}} \right),
\end{equation}
and
\begin{align}
\mathbb{E}\left\{ {\cos y} \right\}  &= \int_{ - \infty }^{ + \infty } {\cos y \cdot f\left( y \right)} dy = \exp \left( { - \overline {{\delta ^2}} } \right)\label{Eq:expected_cosy}\\
\mathbb{E}\left\{ {\sin y} \right\}  &= \int_{ - \infty }^{ + \infty } {\sin y \cdot f\left( y \right)} dy = 0.\label{Eq:expected_siny}
\end{align}
Using the results of (\ref{Eq:expected_cosy}) and (\ref{Eq:expected_siny}), (\ref{Eq:EEP_ExpectedPowerPattern}) is simplified to
\begin{equation}\label{Eq:EEP_ExpectedPowerPattern_Final}
\mathbb{E}\left\{ {{{\left| {{\bf{F}}\left( {\bf{u}} \right)} \right|}^2}} \right\}  = {\left| {{{\bf{F}}_0}\left( {\bf{u}} \right)} \right|^2}{e^{ - \overline {{\delta ^2}} }} + \overline {{\alpha ^2}} \sum\limits_{i = 1}^M {{{\left| {{a_i}} \right|}^2}{{\left| {{{\bf{f}}_i}\left( {\bf{u}} \right)} \right|}^2}}.
\end{equation}
It indicates that the expected power pattern is the exponentially attenuated nominal power pattern, whose attenuation is caused by phase errors of excitation coefficients, plus a ``background" power level proportional to the mean squared error of excitation amplitude.

Plugging (\ref{Eq:EEP_ExpectedPattern_Final}) and (\ref{Eq:EEP_ExpectedPowerPattern_Final}) into (\ref{Eq:NormalizedVariancedefinition}), we obtain
\begin{equation}\label{Eq:rawNormalizedVariance}
\Xi = \overline {{\alpha ^2}} {e^{\overline {{\delta ^2}} }}\frac{{\sum\limits_{i = 1}^M {{{\left| {{a_i}} \right|}^2}{{\left| {{{\bf{f}}_i}\left( {{{\bf{u}}_0}} \right)} \right|}^2}} }}{{{{\left| {{{\bf{F}}_0}\left( {{{\bf{u}}_0}} \right)} \right|}^2}}},
\end{equation}
where the direction is selected in $\bf{u}_0$. Once the control accuracy of the excitation coefficients amplitude and phase is known, the scale factor $\overline {{\alpha ^2}} {e^{\overline {{\delta ^2}} }}$ can be omitted and (\ref{Eq:rawNormalizedVariance}) is reduced to (\ref{Eq:NormalizedVariance}). This concludes the proof of Theorem \ref{theoNormalizedVariance}.
\end{proof}

\subsection{Proof of Theorem \ref{theoConstrainedOpt_a}}\label{proof:theoConstrainedOpt_a}
\begin{proof}
The method of Lagrange multipliers is used to solve this problem. Introducing a scalar multiplier $\Lambda$, one obtains:
\begin{equation}\label{Eq:ConstrainedOpt_L}
L = \frac{{{{\left| {{{\bf{a}}^T}{{\bf{v}}_0}} \right|}^2}}}{{{{\bf{a}}^T}{\bf{B}}{{\bf{a}}^*}}} + \Lambda \frac{{{{\bf{a}}^T}{{\bf{D}}_{{f_0}}}{{\bf{a}}^*}}}{{{{\left| {{{\bf{a}}^T}{{\bf{v}}_0}} \right|}^2}}}.
\end{equation}
By letting the partial derivative of (\ref{Eq:ConstrainedOpt_L}) with respect to $\delta \bf{a}$ equal to zero, one has
\begin{equation}\label{Eq:ConstrainedOpt_VariationL1}
\begin{split}
\delta L &= \left[\frac{{\left( {\delta {{\bf{a}}^T}{{\bf{v}}_0}{{\bf{v}}_0}^H{{\bf{a}}^*} + {{\bf{a}}^T}{{\bf{v}}_0}{{\bf{v}}_0}^H\delta {{\bf{a}}^*}} \right){{\bf{a}}^T}{\bf{B}}{{\bf{a}}^*}}}{{{{\left( {{{\bf{a}}^T}{\bf{B}}{{\bf{a}}^*}} \right)}^2}}} \right.\\
 &\left. \quad - \frac{{{{\left| {{{\bf{a}}^T}{{\bf{v}}_0}} \right|}^2}\left( {\delta {{\bf{a}}^T}{\bf{B}}{{\bf{a}}^*} + {{\bf{a}}^T}{\bf{B}}\delta {{\bf{a}}^*}} \right)}}{{{{\left( {{{\bf{a}}^T}{\bf{B}}{{\bf{a}}^*}} \right)}^2}}} \right]\\
 & \quad + \Lambda \left[\frac{{\left( {\delta {{\bf{a}}^T}{{\bf{D}}_{{f_0}}}{{\bf{a}}^*} + {{\bf{a}}^T}{{\bf{D}}_{{f_0}}}\delta {{\bf{a}}^*}} \right){{\left| {{{\bf{a}}^T}{{\bf{v}}_0}} \right|}^2}}}{{{{\left( {{{\left| {{{\bf{a}}^T}{{\bf{v}}_0}} \right|}^2}} \right)}^2}}} \right.\\
 &\left. \quad - \frac{{{{\bf{a}}^T}{{\bf{D}}_{{f_0}}}{{\bf{a}}^*}\left( {\delta {{\bf{a}}^T}{{\bf{v}}_0}{{\bf{v}}_0}^H{{\bf{a}}^*} + {{\bf{a}}^T}{{\bf{v}}_0}{{\bf{v}}_0}^H\delta {{\bf{a}}^*}} \right)}}{{{{\left( {{{\left| {{{\bf{a}}^T}{{\bf{v}}_0}} \right|}^2}} \right)}^2}}}\right] \\
 & = 0.
\end{split}
\end{equation}
After the rearrangements, we have
\begin{equation}\label{Eq:ConstrainedOpt_VariationL2}
\begin{split}
\delta L &= \delta {{\bf{a}}^T} \left[\frac{{{{\bf{v}}_0}{{\bf{v}}_0}^H{{\bf{a}}^*}{{\bf{a}}^T}{\bf{B}}{{\bf{a}}^*} - {\bf{B}}{{\bf{a}}^*}{{\left| {{{\bf{a}}^T}{{\bf{v}}_0}} \right|}^2}}}{{{{\left( {{{\bf{a}}^T}{\bf{B}}{{\bf{a}}^*}} \right)}^2}}} \right.\\
& \left. \quad + \Lambda \frac{{{{\bf{D}}_{{f_0}}}{{\bf{a}}^*}{{\left| {{{\bf{a}}^T}{{\bf{v}}_0}} \right|}^2} - {{\bf{v}}_0}{{\bf{v}}_0}^H{{\bf{a}}^*}{{\bf{a}}^T}{{\bf{D}}_{{f_0}}}{{\bf{a}}^*}}}{{{{\left( {{{\left| {{{\bf{a}}^T}{{\bf{v}}_0}} \right|}^2}} \right)}^2}}} \right]\\
 &\quad+ \left[\frac{{{{\bf{a}}^T}{\bf{B}}{{\bf{a}}^*}{{\bf{a}}^T}{{\bf{v}}_0}{{\bf{v}}_0}^H - {{\left| {{{\bf{a}}^T}{{\bf{v}}_0}} \right|}^2}{{\bf{a}}^T}{\bf{B}}}}{{{{\left( {{{\bf{a}}^T}{\bf{B}}{{\bf{a}}^*}} \right)}^2}}} \right.\\
 &\left. \quad + \Lambda \frac{{{{\left| {{{\bf{a}}^T}{{\bf{v}}_0}} \right|}^2}{{\bf{a}}^T}{{\bf{D}}_{{f_0}}} - {{\bf{a}}^T}{{\bf{D}}_{{f_0}}}{{\bf{a}}^*}{{\bf{a}}^T}{{\bf{v}}_0}{{\bf{v}}_0}^H}}{{{{\left( {{{\left| {{{\bf{a}}^T}{{\bf{v}}_0}} \right|}^2}} \right)}^2}}}\right]\delta {{\bf{a}}^*} \\
 &= 0.
\end{split}
\end{equation}
Since ${\bf{B}}$ and ${{\bf{D}}_{{f_0}}}$ are Hermitian, Eq. (\ref{Eq:ConstrainedOpt_VariationL2}) can be reduced to
\begin{equation}\label{Eq:ConstrainedOpt_VariationL3}
\begin{split}
\delta L &= 2\Re \left\{ \delta {{\bf{a}}^T}\left[\frac{{{{\bf{v}}_0}{{\bf{v}}_0}^H{{\bf{a}}^*}{{\bf{a}}^T}{\bf{B}}{{\bf{a}}^*} - {\bf{B}}{{\bf{a}}^*}{{\left| {{{\bf{a}}^T}{{\bf{v}}_0}} \right|}^2}}}{{{{\left( {{{\bf{a}}^T}{\bf{B}}{{\bf{a}}^*}} \right)}^2}}}\right.\right.\\
 &\quad \left.\left.+ \Lambda \frac{{{{\bf{D}}_{{f_0}}}{{\bf{a}}^*}{{\left| {{{\bf{a}}^T}{{\bf{v}}_0}} \right|}^2} - {{\bf{v}}_0}{{\bf{v}}_0}^H{{\bf{a}}^*}{{\bf{a}}^T}{{\bf{D}}_{{f_0}}}{{\bf{a}}^*}}}{{{{\left( {{{\left| {{{\bf{a}}^T}{{\bf{v}}_0}} \right|}^2}} \right)}^2}}}\right] \right\} \\
 &= 0.
\end{split}
\end{equation}
Because $\delta {{\bf{a}}^T}$ is arbitrary, Eq. (\ref{Eq:ConstrainedOpt_L}) reaches its extrema when
\begin{multline}\label{Eq:ConstrainedOpt_CriticalPoint1}
\frac{{{{\bf{v}}_0}{{\bf{v}}_0}^H{{\bf{a}}^*}{{\bf{a}}^T}{\bf{B}}{{\bf{a}}^*} - {\bf{B}}{{\bf{a}}^*}{{\left| {{{\bf{a}}^T}{{\bf{v}}_0}} \right|}^2}}}{{{{\left( {{{\bf{a}}^T}{\bf{B}}{{\bf{a}}^*}} \right)}^2}}} \\
+ \Lambda \frac{{{{\bf{D}}_{{f_0}}}{{\bf{a}}^*}{{\left| {{{\bf{a}}^T}{{\bf{v}}_0}} \right|}^2} - {{\bf{v}}_0}{{\bf{v}}_0}^H{{\bf{a}}^*}{{\bf{a}}^T}{{\bf{D}}_{{f_0}}}{{\bf{a}}^*}}}{{{{\left( {{{\left| {{{\bf{a}}^T}{{\bf{v}}_0}} \right|}^2}} \right)}^2}}} = 0.
\end{multline}
By rearranging it, one has
\begin{multline}\label{Eq:ConstrainedOpt_CriticalPoint2}
\left[\frac{{{{\bf{a}}^T}{{\bf{D}}_{{f_0}}}{{\bf{a}}^*}{{\left| {{{\bf{a}}^T}{{\bf{v}}_0}} \right|}^2}}}{{{{\left( {{{\bf{a}}^T}{\bf{B}}{{\bf{a}}^*}} \right)}^2}}}{\bf{B}} + \Lambda  \cdot {\xi^2} \cdot {{\bf{v}}_0}{{\bf{v}}_0}^H \right.\\
- \Lambda  \cdot \xi \cdot {{\bf{D}}_{{f_0}}} \bigg]{{\bf{a}}^*} = \frac{{{{\bf{a}}^T}{{\bf{D}}_{{f_0}}}{{\bf{a}}^*}}}{{{{\bf{a}}^T}{\bf{B}}{{\bf{a}}^*}}}{{\bf{v}}_0}^H{{\bf{a}}^*}{{\bf{v}}_0}.
\end{multline}
Since ${{\bf{v}}_0}^H{{\bf{a}}^*} \ne 0$, Eq. (\ref{Eq:ConstrainedOpt_a}) can be derived. This completes the proof of Theorem \ref{theoConstrainedOpt_a}.
\end{proof}

\subsection{Proof of Theorem \ref{theoConstrainedOpt_p}}\label{proof:theoConstrainedOpt_p}
\begin{proof}
It is rather difficult to directly solve (\ref{Eq:ConstrainedOpt_EqOfp}) since it involves the inversion of the matrix $\bf{K}$ which contains the unknown $p$. To circumvent this obstacle, we look at (\ref{Eq:ConstrainedOpt_EqOfp}) from another perspective, rewriting it as
\begin{equation}\label{Eq:ConstrainedOpt_EqOfp_InnerProduct}
\left( {{{\bf{v}}_0},{{\bf{K}}^{ - 1}}\left( {\xi  \cdot {{\bf{v}}_0}{{\bf{v}}_0}^H - {{\bf{D}}_{{f_0}}}} \right){{\bf{K}}^{ - 1}}{{\bf{v}}_0}} \right) = 0,
\end{equation}
which means the vector ${{\bf{K}}^{ - 1}}\left( {\xi  \cdot {{\bf{v}}_0}{{\bf{v}}_0}^H - {{\bf{D}}_{{f_0}}}} \right){{\bf{K}}^{ - 1}}{{\bf{v}}_0}$ lies in the orthogonal complement of ${{\bf{v}}_0}$, i.e., $\text{span}\{ {{{\bf{v}}_2},{{\bf{v}}_3}, \cdots ,{{\bf{v}}_M}} \}$.

Once ${{\bf{v}}_0}$ is given, vectors ${{{\bf{v}}_2},{{\bf{v}}_3}, \cdots ,{{\bf{v}}_M}}$ can be chosen by SVD. Obviously, ${{\bf{K}}^{ - 1}}\left( {\xi  \cdot {{\bf{v}}_0}{{\bf{v}}_0}^H - {{\bf{D}}_{{f_0}}}} \right){{\bf{K}}^{ - 1}}{{\bf{v}}_0}$ must be a linear combination of $\{ {{{\bf{v}}_2},{{\bf{v}}_3}, \cdots ,{{\bf{v}}_M}} \}$:
\begin{equation}\label{Eq:ConstrainedOpt_LinearComb}
{{\bf{K}}^{ - 1}}\left( {\xi  \cdot {{\bf{v}}_0}{{\bf{v}}_0}^H - {{\bf{D}}_{{f_0}}}} \right){{\bf{K}}^{ - 1}}{{\bf{v}}_0} = \sum\limits_{i = 2}^M {{h_i}{{\bf{v}}_i}} , 
\end{equation}
where $\{ h_i \}$ is a set of constant scalars. Further, this equation can be reorganized as:
\begin{equation}\label{Eq:ConstrainedOpt_WH=0}
{\bf{W}\bf{h}} = \bf{0},
\end{equation}
where $\bf{h}$ is a vector defined as 
\begin{equation}\label{Eq:ConstrainedOpt_H}
{\bf{h}} = \begin{bmatrix}
-1  & h_2  & h_3 & \cdots  & h_M
\end{bmatrix}^T.
\end{equation}
Since $\bf{h}$ is not a zero vector, the determinant of $\bf{W}$ must be zero according to Cramer's Rule. This proves Therorem \ref{theoConstrainedOpt_p}.
\end{proof}

%%%%%%%%%%%%%%%%%%%%%%%%%%%%%%%%%%%%% 
\bibliographystyle{IEEEtran}
\bibliography{bib/allCitations}

\end{document}